\documentclass[aps,pre,amsmath,twocolumn,10pt,superscriptaddress,longbibliography]{revtex4-1}
\usepackage[dvipdfmx]{graphicx}
\usepackage[dvipdfmx,colorlinks=true]{hyperref}
\usepackage{color}
\usepackage{bm}
\usepackage{braket}
\usepackage{amsmath,amssymb}
\usepackage{mathtools}

\usepackage{physics}
\usepackage[normalem]{ulem}

\begin{document}
\title{
Scalable construction of force-constant basis sets for large-scale anharmonic lattice-dynamical calculations}
\author{Atsuto \surname{Seko}}
\email{seko@cms.mtl.kyoto-u.ac.jp}
\affiliation{Department of Materials Science and Engineering, Kyoto University, Kyoto 606-8501, Japan}
\author{Atsushi \surname{Togo}}
\affiliation{Center for Basic Research on Materials National Institute for Materials Science, Tsukuba, Ibaraki 305-0047, Japan}

\date{\today}

\begin{abstract}
A projector-based formulation of force constants in crystalline materials provides a systematic framework for constructing force-constant bases and determining force constants from force--displacement datasets while rigorously satisfying crystal symmetry, permutation symmetry, and translational invariance.
In this work, we develop an efficient eigenvalue solver for projection matrices and integrate it into the projector-based framework.
The proposed method substantially reduces the computational cost and enables practical calculations for large-scale systems, low-symmetry crystals, and higher-order force constants that are difficult to treat using conventional approaches.
Applications to self-consistent phonon calculations for assessing the grain-boundary excess free energy, lattice thermal conductivity calculations in complex compounds, and fourth-order force-constant estimations demonstrate the efficiency and robustness of the proposed framework for large and complex materials systems.
\end{abstract}

\maketitle

\section{Introduction}
\label{symfc2:sec-introduction}

Force constants in crystalline materials play a central role in the calculation of phonon-related properties, including lattice dynamics, thermal transport, and vibrational thermodynamics \cite{IntroductionToLatticeDynamics, Thermodynamics-of-crystals, Ziman-electrons-phonons}.
A common strategy for determining force constants is to employ the supercell appoach combined with finite atomic displacements \cite{Parlinski-phonon-1997, Hellman-TDEP-2013, Tadano-ALM-2018, hiPhive, phonopy-phono3py-JPSJ}. 
In this framework, systematic or random atomic displacements are introduced into supercells, and the resulting atomic forces are calculated using density-functional theory (DFT). 
The supercell force constants are then estimated from force--displacement datasets using least-squares fitting or numerical differentiation techniques.
Similarly, self-consistent phonon methods determine effective supercell force constants from structures representing finite-temperature atomic distributions \cite{Errea-SSCHA-2013, van-Roekeghem-2020}.
In these methods, the force constants are obtained either through free-energy minimization or by fitting the force--displacement dataset derived from sampled structures.

Owing to the symmetry properties inherent in force constants, the number of independent force-constants can be substantially reduced by imposing invariance under crystal space-group operations, permutations of force-constant indices, and infinitesimal translations of the crystal \cite{Thermodynamics-of-crystals, Tadano-ALM-2018, hiPhive, Physical-Properties-of-Crystals, el-batanouny_wooten_2008}. 
A complete orthonormal basis set for the force constants enables the accurate determination of force constants while rigorously satisfying these invariance conditions \cite{Dynamics-of-perfect-crystals}.
Various methods, including approaches based on compressed sensing, have been developed to estimate force constants with high accuracy \cite{Zhou-PRL-compressive-sensing-FC-2014, Tadano-2015}. 
Nevertheless, despite the availability of several efficient approaches \cite{Tadano-ALM-2018, hiPhive, Hellman-TDEP-2013}, both force-constant estimation and the construction of force-constant basis sets remain computationally demanding, particularly for large-scale systems and higher-order force constants.

Recently, the authors proposed a projector-based approach for constructing complete force-constant basis sets and efficiently estimating force constants while rigorously satisfying the required invariance conditions \cite{PhysRevB.110.214302}. 
As described in Sec.~\ref{symfc2:sec-methodology}, the projector-based framework implemented in the \textsc{symfc} code \cite{symfc-project} incorporates several techniques for reducing the projection matrices to be solved, thereby substantially lowering the computational cost of both force-constant basis-set construction and force-constant estimation.
With these developments, systematic calculations of lattice thermal conductivity (LTC) based on third-order force constants \cite{10.1063/5.0211296}, as well as self-consistent phonon free-energy calculations for large numbers of structures involving many independent force-constant components \cite{wakai2026globalstructuresearchesvarying}, have become computationally feasible.

However, the construction of complete force-constant basis sets and the estimation of force constants can still be computationally demanding. 
For example, self-consistent phonon calculations with second-order effective force constants become expensive for systems requiring large supercells or for low-symmetry crystals, where only a limited number of space-group operations can be exploited. 
When higher-order force constants, such as third- and fourth-order terms, are considered, the basis-set construction becomes computationally prohibitive \cite{PhysRevB.110.214302}.
As discussed in more detail in Sec.~\ref{symfc2:sec-solving-eigenvalue-problem-projector}, the primary computational bottleneck of the previous approach is solving the eigenvalue problems associated with the reduced, yet still large, dense projection matrices that satisfy all invariance conditions.

In this study, we develop an efficient solver for the eigenvalue problems associated with projection matrices and their products, and incorporate it into the projector-based framework for force-constant basis-set construction.
This improvement substantially reduces the computational cost of constructing force-constant basis sets and significantly extends the applicability of the projector-based approach.
In particular, it enables the practical treatment of large-scale systems, low-symmetry crystals, and higher-order force constants that are challenging to handle using conventional methods. 
To demonstrate the effectiveness of the proposed method, we present several representative applications, including self-consistent phonon calculations for large systems, systematic construction of third-order force-constant basis sets for a wide range of compounds, lattice thermal conductivity calculations for low-symmetry ternary and quaternary compounds, and fourth-order force-constant estimation.

This paper is organized as follows. 
In Sec.~\ref{symfc2:sec-methodology}, we summarize the formulation of the projection matrices for force constants introduced in the previous study. 
In Sec.~\ref{symfc2:sec-solving-eigenvalue-problem-projector}, we present the fundamental concepts underlying the proposed method and describe a practical recursive algorithm for constructing complete force-constant basis sets with improved computational efficiency.
We also clarify the origin of the computational bottleneck in the projector-based force-constant basis-set construction proposed in the previous study.
Section~\ref{symfc2:sec-results} presents several representative applications of the proposed method. 
Finally, Sec.~\ref{symfc2:sec-conclusion} concludes the paper.

\section{Force constant projector}
\label{symfc2:sec-methodology}

\subsection{Force constants in crystals}
\label{symfc2:sec-force-constants}

Force constants of a crystal are derived from the Taylor expansion of the potential energy $\mathcal{V}$ with respect to atomic displacements $\{u_{i\alpha}\}$.
When considering a supercell containing $N$ atoms, the potential energy can be expressed as
\begin{eqnarray}
\mathcal{V} &=& \Theta_0 + \sum_{i\alpha} \Theta_{i\alpha} u_{i\alpha}
+ \frac{1}{2} \sum_{i\alpha,j\beta} \Theta_{i\alpha,j\beta} u_{i\alpha} u_{j\beta} \nonumber \\
&+& \frac{1}{3!} \sum_{i\alpha,j\beta,k\gamma} \Theta_{i\alpha,j\beta,k\gamma} 
u_{i\alpha} u_{j\beta} u_{k\gamma} \nonumber \\
&+& \frac{1}{4!} \sum_{i\alpha,j\beta,k\gamma,l\delta} \Theta_{i\alpha,j\beta,k\gamma,l\delta}
u_{i\alpha} u_{j\beta} u_{k\gamma} u_{l\delta} \nonumber \\
&+& \cdots,
\end{eqnarray}
where $i$, $j$, $k$, and $l$ denote atomic indices within the supercell, and $\alpha$, $\beta$, $\gamma$, and $\delta$ represent Cartesian components.
The second-, third-, and fourth-order supercell force constants are denoted by $\Theta_{i\alpha,j\beta}$, $\Theta_{i\alpha,j\beta,k\gamma}$, and $\Theta_{i\alpha,j\beta,k\gamma,l\delta}$, respectively.

In formulating the force-constant projection matrices, it is convenient to represent the force constants as a column vector, $\bm{\theta}$, obtained by flattening the force constants $\Theta_{i\alpha,j\beta}$, $\Theta_{i\alpha,j\beta,k\gamma}$, and $\Theta_{i\alpha,j\beta,k\gamma,l\delta}$.
The second-order force constants are represented as a column vector with $9N^2$ elements, denoted as
\begin{equation}
\bm{\theta}^{\rm (FC2)} = [\Theta_{1x,1x}, \Theta_{1x,1y}, \Theta_{1x,1z}, \Theta_{1x,2x}, \cdots]^\top,
\end{equation}
where the components \(\Theta_{i\alpha,j\beta}\) are listed in ascending order of the indices \((i, \alpha, j, \beta)\).
The third- and fourth-order force constants are likewise represented as column vectors containing \(27N^3\) and \(81N^4\) elements, respectively, with the elements arranged in ascending order of the composite indices \((i,\alpha,j,\beta,k,\gamma)\) and \((i,\alpha,j,\beta,k,\gamma,l,\delta)\).

\subsection{Force-constant projector and basis set}
\label{symfc2:sec-force-constant-projector-and-basis-set}

We now define an orthogonal projection matrix to derive force constants that are invariant under crystal space-group operations, index permutations, and infinitesimal translations.
We refer to the orthogonal projection matrix that maps the force constants onto the subspace satisfying all these invariance conditions as the full projection matrix, denoted by $\bm{P}$.
Accordingly, when the full projection matrix is applied to an arbitrary set of force constants, $\bm{\theta}'$, as
\begin{equation}
\label{symfc2:eq-projection-apply}
\bm{\theta} = \bm{P}\bm{\theta}',
\end{equation}
the resulting force constants $\bm{\theta}$ satisfy all the invariance conditions.

The full projection matrix can be represented by the eigendecomposition
\begin{equation}
\label{symfc2:eq-full-projector-eigen}
\bm{P} = \bm{B}\bm{B}^\top,
\end{equation}
where $\bm{B}$ is the matrix whose columns are the eigenvectors of the projection matrix $\bm{P}$ corresponding to the unit eigenvalue.
These eigenvectors form a complete basis set for the force constants satisfying all the invariance conditions.
Conversely, given such a complete basis set, the full projection matrix can be constructed as Eq. (\ref{symfc2:eq-full-projector-eigen}).

Any set of force constants satisfying all the invariance conditions can be expressed as a linear combination of the force-constant basis vectors:
\begin{equation}
\label{symfc2:eq-fc-expansion}
\bm{\theta} = \bm{B}\bm{c},
\end{equation}
where $\bm{c}$ is the vector of expansion coefficients.
This representation is useful for estimating the force constants from force--displacement datasets using least-squares methods and for optimizing the force constants based on variational principles, as employed in self-consistent phonon calculations.

\subsection{Construction of the force-constant basis set}
\subsubsection{Difficulty in basis-set construction}
\label{symfc2:sec-difficulty-in-basis-set-construction}

Although the concepts of the full projection matrix and the complete set of force-constant basis vectors are straightforward, constructing the full projector and basis set is computationally demanding.
In particular, the computational cost becomes prohibitive for systems requiring a large number of basis vectors, such as higher-order force constants, systems containing many atoms, or systems with few space-group operations.
One simple reason for this is that the number of force-constant components grows rapidly with both the order of the force constants and the number of atoms.
When the number of atoms in a supercell is denoted by \(N\), the numbers of elements in the second-, third-, and fourth-order force constants are \(9N^2\), \(27N^3\), and \(81N^4\), respectively.
Another reason is the noncommutative nature of the orthogonal projection matrices associated with index permutations and the translational sum rules.
This noncommutativity makes the problem considerably more complex, and therefore we discuss it in more detail below.

Let \(\bm{P}_{\mathrm{spg}}\), \(\bm{P}_{\mathrm{perm}}\), and \(\bm{P}_{\mathrm{trsum}}\) denote the orthogonal projection matrices associated with space-group operations, index permutations, and translational sum rules, respectively.
Their mathematical definitions are provided in Ref.~\cite{PhysRevB.110.214302}.
Applying these projection matrices to the force constants \(\bm{\theta}\), i.e., \(\bm{P}_{\mathrm{spg}}\bm{\theta}\), \(\bm{P}_{\mathrm{perm}}\bm{\theta}\), and \(\bm{P}_{\mathrm{trsum}}\bm{\theta}\), yields force constants that satisfy the corresponding invariance conditions under space-group operations, index permutations, and infinitesimal translations, respectively.
Although applying each projection matrix individually is straightforward, the noncommutativity of \(\bm{P}_{\mathrm{perm}}\) and \(\bm{P}_{\mathrm{trsum}}\), i.e., \([\bm{P}_{\mathrm{perm}}, \bm{P}_{\mathrm{trsum}}] \neq 0\), prevents them from being applied sequentially to obtain force constants that exactly satisfy all invariance conditions simultaneously.

As a consequence, the full projection matrix cannot be written as the simple product of the individual projection matrices \cite{Neumann1949OnRO,halperin1962product,PIZIAK199967,galantai2013projectors}, i.e.,
\begin{equation}
\bm{P} \neq \bm{P}_{\mathrm{spg}} \bm{P}_{\mathrm{perm}} \bm{P}_{\mathrm{trsum}}.
\end{equation}
Furthermore, when force constants that satisfy permutational symmetry, \(\bm{P}_{\mathrm{perm}}\bm{\theta}\), are further projected by \(\bm{P}_{\mathrm{trsum}}\), the resulting force constants no longer satisfy the permutational invariance condition, which can be expressed as
\begin{equation}
\bm{P}_{\mathrm{trsum}} (\bm{P}_{\mathrm{perm}} \bm{\theta}) \notin \mathrm{Span} (\bm{B}_{\mathrm{perm}}),
\end{equation}
where \(\bm{B}_{\mathrm{perm}}\) denotes the complete basis set associated with \(\bm{P}_{\mathrm{perm}}\), satisfying \(\bm{P}_{\mathrm{perm}} \bm{B}_{\mathrm{perm}} = \bm{B}_{\mathrm{perm}}\).

\subsubsection{Reformulation of the eigenvalue problem for the projection matrix}
\label{symfc2:sec-reformulation-eigenvalue-problem}

Despite the aforementioned difficulties in constructing complete basis sets that satisfy all invariance conditions, the full projection matrix or basis set can be obtained from a set of individual projection matrices.
As shown in Ref.~\cite{PhysRevB.110.214302}, the full projection matrix enforcing all invariance conditions is given by
\cite{Neumann1949OnRO,halperin1962product,PIZIAK199967,galantai2013projectors}
\begin{equation}
\label{symfc2:eq-Neumann-Halperin}
\bm{P} = \lim_{p \to \infty} \bm{P}_{\mathrm{spg}} \bm{P}_{\mathrm{perm}}
(\bm{P}_{\mathrm{trsum}} \bm{P}_{\mathrm{perm}})^p.
\end{equation}
Moreover, the complete force-constant basis set of $\bm{P}$, denoted by $\bm{B}$, consists of the common eigenvectors of $\bm{P}$, $\bm{P}_{\mathrm{spg}}$, $\bm{P}_{\mathrm{perm}}$, and $\bm{P}_{\mathrm{trsum}}$
with eigenvalue unity, satisfying
\begin{equation}
\label{symfc2:eq-spg-perm-trsum-eigenequation}
\bm{P}_{\mathrm{spg}} \bm{B} = \bm{B},
\:\: \bm{P}_{\mathrm{perm}} \bm{B} = \bm{B},
\:\: \bm{P}_{\mathrm{trsum}} \bm{B} = \bm{B}.
\end{equation}
This formulation properly accounts for the commutation relations among the projection matrices:
\begin{equation}
\label{symfc2:eq-commutation}
[ \bm{P}_{\rm spg}, \bm{P}_{\rm perm}  ]  =  0, 
\:\: \left[ \bm{P}_{\rm spg}, \bm{P}_{\rm trsum} \right]  =  0,
\:\: \left[ \bm{P}_{\rm perm}, \bm{P}_{\rm trsum} \right] \neq  0.
\end{equation}
Although Eq.~(\ref{symfc2:eq-Neumann-Halperin}) involves the limit of infinitely many matrix products, the complete basis set of the full projection matrix can nevertheless be obtained by solving the eigenvalue problem for the matrix $\bm{M}_{p=1}$, defined as \cite{PhysRevB.110.214302}
\begin{eqnarray}
\label{symfc2:eq-approx-projector}
\bm{M}_{p=1}
&=&
\bm{P}_{\rm spg} \bm{P}_{\rm perm} \bm{P}_{\rm trsum} \bm{P}_{\rm perm} \nonumber \\
&=&
\bm{P}_{\rm spg} \bm{P}_{\rm perm} \bm{P}_{\rm trsum} \bm{P}_{\rm spg} \bm{P}_{\rm perm},
\end{eqnarray}
where the second equality follows from the commutation relations among the projection matrices given by Eq.~(\ref{symfc2:eq-commutation}) and the idempotency of $\bm{P}_{\mathrm{spg}}$, i.e., $\bm{P}_{\mathrm{spg}}^2=\bm{P}_{\mathrm{spg}}$.
The matrix \(\bm{M}_{p=1}\) is not itself a projection matrix, and its eigenvalues range from 0 to 1.
Therefore, its eigendecomposition is given by
\begin{equation}
\label{symfc2:eq-eigendecomposition}
\bm{M}_{p=1} = \bm{B} \bm{B}^\top + \bm{C} \bm{\Lambda}_c \bm{C}^\top,
\end{equation}
where \(\bm{B}\) and \(\bm{C}\) denote the eigenvectors associated with eigenvalues equal to one and those strictly less than one (but non-negative), respectively.
Here, \(\bm{\Lambda}_c\) is a diagonal matrix whose elements are non-negative eigenvalues strictly less than one.
Only the eigenvectors corresponding to the unit eigenvalue, namely \(\bm{B}\), form a complete basis satisfying all invariance conditions.
Accordingly, the full projection matrix can be constructed from \(\bm{B}\) obtained from \(\bm{M}_{p=1}\) as
\begin{equation}
\label{symfc2:eq-full-projector}
\bm{P} = \bm{B} \bm{B}^\top.
\end{equation}

\subsubsection{Reduction techniques}
\label{symfc2:sec-reduction-techniques}

We then provide an overview of the procedure for reducing the dimension of the eigenvalue problem for \(\bm{M}_{p=1}\), which enables its efficient solution \cite{PhysRevB.110.214302}.
First, the eigenvalue problem for the product \(\bm{P}_{\mathrm{spg}} \bm{P}_{\mathrm{perm}}\) is solved.
Because the eigenvectors of this product can be chosen to be highly sparse, the eigenvalue problem can be solved efficiently in terms of both computational time and memory usage.
Moreover, because \(\bm{P}_{\mathrm{spg}}\) and \(\bm{P}_{\mathrm{perm}}\) commute, their product is itself a projection matrix.
Consequently, \(\bm{P}_{\mathrm{spg}} \bm{P}_{\mathrm{perm}}\) can be written as
\begin{equation}
\label{symfc2:eq-spg-perm-eigen}
\bm{P}_{\mathrm{spg}} \bm{P}_{\mathrm{perm}}
=
\bm{B}_{\mathrm{spg} \cap \mathrm{perm}}
\bm{B}_{\mathrm{spg} \cap \mathrm{perm}}^\top,
\end{equation}
where \(\bm{B}_{\mathrm{spg} \cap \mathrm{perm}}\) denotes a complete basis set satisfying both space-group and permutational invariance.

Since the number of columns of $\bm{B}_{\rm spg \cap \rm perm}$ is much smaller than the dimension of the full projection matrix, it can be used to reduce the eigenvalue problems for $\bm{P}_{\rm trsum}$ and $\bm{M}_{p=1}$ to the subspace spanned by $\bm{B}_{\rm spg \cap \rm perm}$.
By substituting Eq.~(\ref{symfc2:eq-spg-perm-eigen}) into Eq.~(\ref{symfc2:eq-approx-projector}), $\bm{M}_{p=1}$ can be written as
\begin{eqnarray}
\label{symfc2:eq-trsum-compress}
\bm{M}_{p=1}
&=&
\bm{B}_{\rm spg \cap \rm perm}
(\bm{B}_{\rm spg \cap \rm perm}^\top
\bm{P}_{\rm trsum}
\bm{B}_{\rm spg \cap \rm perm})
\bm{B}_{\rm spg \cap \rm perm}^\top
\nonumber \\
&=&
\bm{B}_{\rm spg \cap \rm perm}
\bm{M}_{\rm trsum}'
\bm{B}_{\rm spg \cap \rm perm}^\top,
\end{eqnarray}
where
\begin{equation}
\label{symfc2:eq-trsum-compress-core}
\bm{M}_{\rm trsum}'
=
\bm{B}_{\rm spg \cap \rm perm}^\top
\bm{P}_{\rm trsum}
\bm{B}_{\rm spg \cap \rm perm}
\end{equation}
is the representation of $\bm{P}_{\rm trsum}$ in the basis
$\bm{B}_{\rm spg \cap \rm perm}$.
Therefore, the eigendecomposition of $\bm{M}_{p=1}$ reduces to that of $\bm{M}_{\rm trsum}'$.
Once the eigendecomposition of $\bm{M}_{\rm trsum}'$ is obtained as
\begin{equation}
\label{symfc2:eq-trsum-projector}
\bm{M}_{\rm trsum}'
=
\bm{B}_{\rm trsum}'
\bm{B}_{\rm trsum}'^\top
+
\bm{C}_{\rm trsum}'
\bm{\Lambda}_{c'}
\bm{C}_{\rm trsum}'^\top,
\end{equation}
the complete basis set $\bm{B}$ for the full projection matrix $\bm{P}$ is recovered as
\begin{equation}
\bm{B}
=
\bm{B}_{\rm spg \cap \rm perm}
\bm{B}_{\rm trsum}'.
\end{equation}

\section{Solving the Eigenvalue Problem for Projection Matrices}
\label{symfc2:sec-solving-eigenvalue-problem-projector}

\subsection{Strategies for handling large-scale eigenvalue problems}

As described in Sec. \ref{symfc2:sec-reduction-techniques}, the eigenvalue problems arising in the construction of a complete force-constant basis set can be reduced in dimension.
When the resulting reduced eigenvalue problems are sufficiently small to be solved practically using standard eigensolvers, it is sufficient to solve them directly.
On the other hand, when the reduced eigenvalue problems remain prohibitively large, additional strategies are required to make their solution computationally feasible.

The first strategy is to exploit the block-diagonal structure of the projection matrices, following the procedure described in Sec. \ref{symfc2:sec-block-diagonal-structure}, which was introduced in our previous study \cite{PhysRevB.110.214302}.
If the resulting eigenvalue problems are still prohibitively large after applying this strategy, the procedures proposed in this study can be employed.
These procedures are described in Secs.~\ref{symfc2:sec-projection-matrix-partitioning} and \ref{symfc2:sec-recursive-algorithm}.

For the eigenvalue problem associated with $\bm{M}_{\rm trsum}'$ defined in Eq.~(\ref{symfc2:eq-trsum-compress-core}), solving the eigenvalue problem remains a significant computational bottleneck in the overall procedure for constructing force-constant basis sets of large-scale systems.
This is because $\bm{M}_{\rm trsum}'$ is generally not sufficiently sparse, and it cannot be decomposed into block-diagonal form with sufficiently small block submatrices. 
Therefore, the procedures described in Secs.~\ref{symfc2:sec-projection-matrix-partitioning} and \ref{symfc2:sec-recursive-algorithm} are particularly useful for this eigenvalue problem.

\subsection{Use of block-diagonal structure}
\label{symfc2:sec-block-diagonal-structure}

An effective strategy for solving such eigenvalue problems is to exploit their block-diagonal structure \cite{PhysRevB.110.214302}.
The block-diagonal form of the orthogonal projection matrix $\bm{P}$ is given by
\begin{equation}
\bm{P} =
\begin{bmatrix}
\bm{S}_{11} & \bm{0} & \bm{0} \\
\bm{0} & \bm{S}_{22} & \bm{0} \\
\bm{0} & \bm{0} & \bm{S}_{33}
\end{bmatrix},
\end{equation}
where appropriate permutations of rows and columns have been applied.
The submatrices $\bm{S}_{11}$, $\bm{S}_{22}$, and $\bm{S}_{33}$ are themselves orthogonal projection matrices, which ensures that $\bm{P}$ is also an orthogonal projection matrix.

When solving the eigenvalue problems for these smaller submatrices and obtaining the eigenvectors of $\bm{S}_{11}$, $\bm{S}_{22}$, and $\bm{S}_{33}$ as
\begin{equation}
\bm{S}_{11}\bm{b}_{11} = \bm{b}_{11},
\:
\bm{S}_{22}\bm{b}_{22} = \bm{b}_{22}, 
\: 
\bm{S}_{33}\bm{b}_{33} = \bm{b}_{33},
\end{equation}
the vectors 
\begin{equation}
\bm{b} = 
\begin{bmatrix}
\bm{b}_{11} \\
\bm{0}   \\
\bm{0}   \\
\end{bmatrix},
\begin{bmatrix}
\bm{0} \\
\bm{b}_{22}   \\
\bm{0}   \\
\end{bmatrix},
\begin{bmatrix}
\bm{0} \\
\bm{0}   \\
\bm{b}_{33}   \\
\end{bmatrix}
\end{equation}
can be selected as eigenvectors of $\bm{P}$.
Thus, the eigenvectors of the original projection matrix $\bm{P}$ can be obtained by solving the eigenvalue problems of the smaller matrices.

\subsection{Projetion matrix partitioning}
\label{symfc2:sec-projection-matrix-partitioning}

To address eigenvalue problems for large matrices, this study proposes an efficient algorithm applicable to orthogonal projection matrices and products of orthogonal projection matrices.
The proposed algorithm is based on partitioning a projection matrix into submatrices and solving the eigenvalue problems associated with these submatrices.
The algorithm is applicable to arbitrary orthogonal projection matrices and their products arising in more general contexts beyond force-constant basis-set construction.

This subsection presents the fundamental idea underlying the proposed algorithm.
The discussion in this subsection focuses on partitioning a projection matrix into $2 \times 2$ submatrices.
However, the discussion can be straightforwardly extended to the partitioning of a projection matrix into $n \times n$ submatrices with $n>2$.

We consider the partitioning of the orthogonal projection matrix $\bm{P}$ into $2 \times 2$ submatrices:
\begin{equation}
\bm{P} =
\begin{bmatrix}
\bm{S}_{11}      & \bm{S}_{12} \\
\bm{S}_{12}^\top & \bm{S}_{22}
\end{bmatrix},
\end{equation}
where $\bm{S}_{11}$ and $\bm{S}_{22}$ are square matrices referred to as principal submatrices.
When the principal submatrices $\bm{S}_{11}$ and $\bm{S}_{22}$ have eigenvectors corresponding to the eigenvalue 1, we denote them by $\bm{B}_{11}$ and $\bm{B}_{22}$, respectively:
\begin{eqnarray}
\bm{S}_{11} \bm{B}_{11} &=& \bm{B}_{11}, \nonumber \\
\bm{S}_{22} \bm{B}_{22} &=& \bm{B}_{22}.
\end{eqnarray}
Meanwhile, these principal submatrices can also be expressed in terms of products involving orthogonal projection matrices as
\begin{equation}
\bm{P}_1 \bm{P} \bm{P}_1 =
\begin{bmatrix}
\bm{S}_{11} & \bm{0} \\
\bm{0} & \bm{0}
\end{bmatrix}
,\:\:
\bm{P}_2 \bm{P} \bm{P}_2 =
\begin{bmatrix}
\bm{0} & \bm{0} \\
\bm{0} & \bm{S}_{22}
\end{bmatrix},
\end{equation}
where $\bm{P}_1$ and $\bm{P}_2$ are projection matrices defined using the identity matrices $\bm{I}_{1}$ and $\bm{I}_{2}$, whose dimensions are equal to those of $\bm{S}_{11}$ and $\bm{S}_{22}$, respectively. 
They are given by
\begin{equation}
\label{symfc2:eq-p1matrix}
\bm{P}_1 =
\begin{bmatrix}
\bm{I}_{1} & \bm{0} \\
\bm{0} & \bm{0}
\end{bmatrix}
,\:\:
\bm{P}_2 =
\begin{bmatrix}
\bm{0} & \bm{0} \\
\bm{0} & \bm{I}_2
\end{bmatrix}.
\end{equation}
Thus, since $\bm{B}_{11}$ and $\bm{B}_{22}$ are obtained by solving the eigenvalue problems for the products of orthogonal projection matrices $\bm{P}_1 \bm{P} \bm{P}_1$ and $\bm{P}_2 \bm{P} \bm{P}_2$, respectively, the matrix $\bm{B}^{(1)}$,
\begin{equation}
\label{symfc2:eq-block-eigenvectors1}
\bm{B}^{(1)} =
\begin{bmatrix}
\bm{B}_{11} & \bm{0} \\
\bm{0}      & \bm{B}_{22}
\end{bmatrix},
\end{equation}
consists of a set of mutually independent eigenvectors of the original matrix $\bm{P}$, following properties analogous to those described in Sec. \ref{symfc2:sec-reformulation-eigenvalue-problem}.

This partitioning-based procedure is effective when many eigenvectors corresponding to the eigenvalue 1 can be obtained by solving the eigenvalue problems for the principal submatrices.
For $\bm{M}_{\rm trsum}'$, a practical partitioning that yields submatrices of reasonable size can generally identify a large number of eigenvectors corresponding to the eigenvalue 1.

All remaining eigenvectors must be identified from the complementary projection matrix
$\bm{P}^{(2)} = \bm{P} - \bm{B}^{(1)}\bm{B}^{(1)\top}$ to construct a complete basis set for $\bm{P}$.
By solving the eigenvalue problem for $\bm{P}^{(2)}$, the eigenvectors corresponding to the eigenvalue 1 are obtained as
\begin{equation}
\bm{P}^{(2)} \bm{B}^{(2)} = \bm{B}^{(2)}.
\end{equation}
Finally, the complete basis set for $\bm{P}$ is given by the column-wise concatenation
\begin{equation}
\bm{B} = [\bm{B}^{(1)} : \bm{B}^{(2)}].
\end{equation}
In practice, an additional efficient procedure is employed to solve the eigenvalue problem for $\bm{P}^{(2)}$, since $\bm{P}^{(2)}$ can become large.
Details of this procedure are provided in Appendix~\ref{symfc2:appendix-complementary-projector}.

Note that, although the fundamental concept for obtaining eigenvectors of an orthogonal projection matrix has been introduced, the same approach can also be extended to products of noncommutative orthogonal projection matrices.
A proof for this case is provided in Appendix~\ref{symfc2:appendix-proof2}.
In this case, eigenvectors associated with non-negative eigenvalues smaller than one remain after solving the eigenvalue problem for $\bm{P}^{(2)}$.
These eigenvectors can be used to define the complementary subspace of the vector space spanned by the columns of $\bm{B} = [\bm{B}^{(1)} : \bm{B}^{(2)}]$.

\subsection{Recursive algorithm}
\label{symfc2:sec-recursive-algorithm}

\begin{figure}[tbp]
\includegraphics[clip,width=\linewidth]{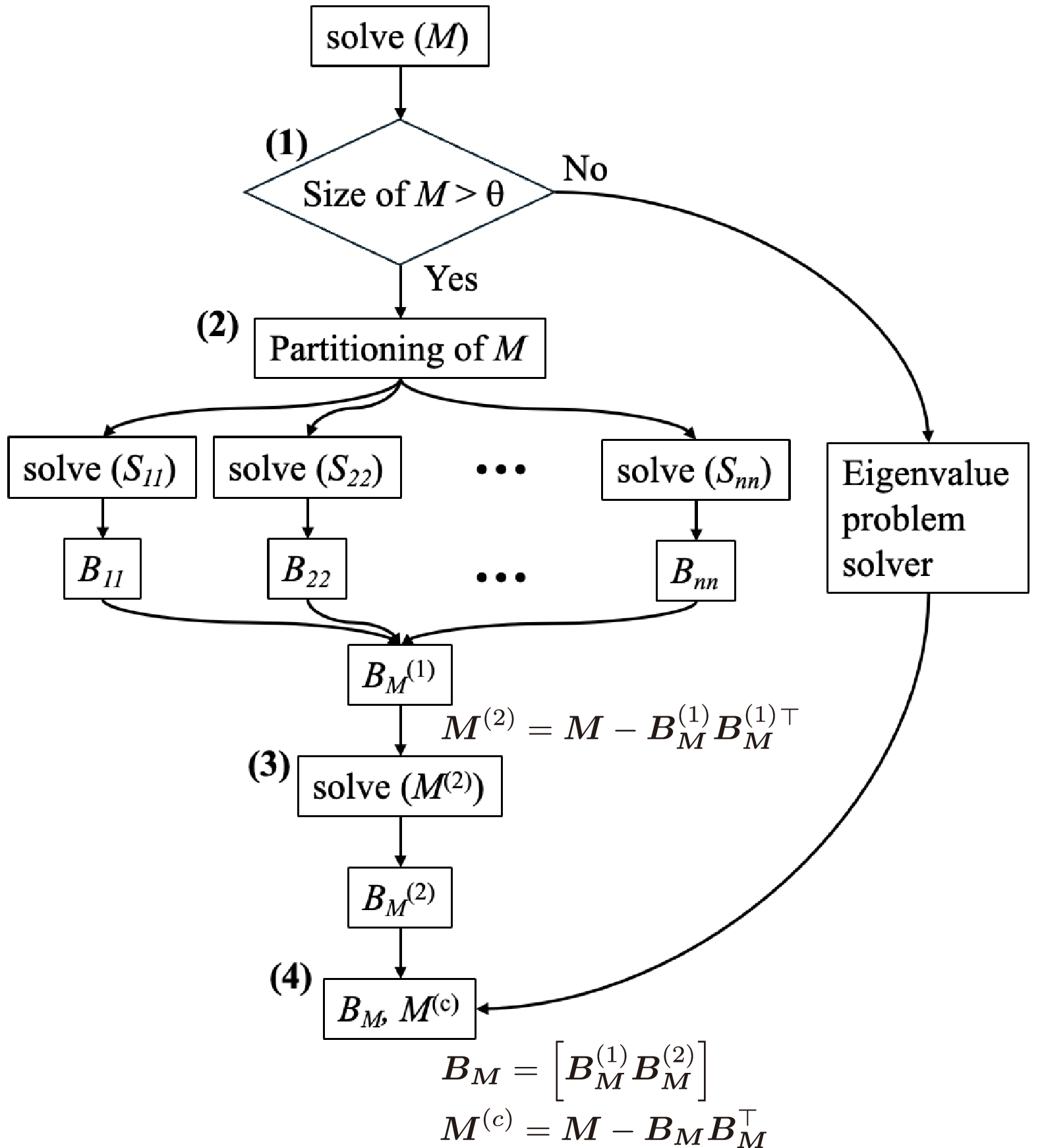}
\caption{
Flowchart of the current recursive algorithm for solving the eigenvalue problem of the matrix $\bm{M}$.
Here, $\theta$ denotes the threshold value used for submatrix partitioning.
}
\label{symfc2:Fig-flowchart}
\end{figure}

\begin{figure}[tbp]
\includegraphics[clip,width=\linewidth]{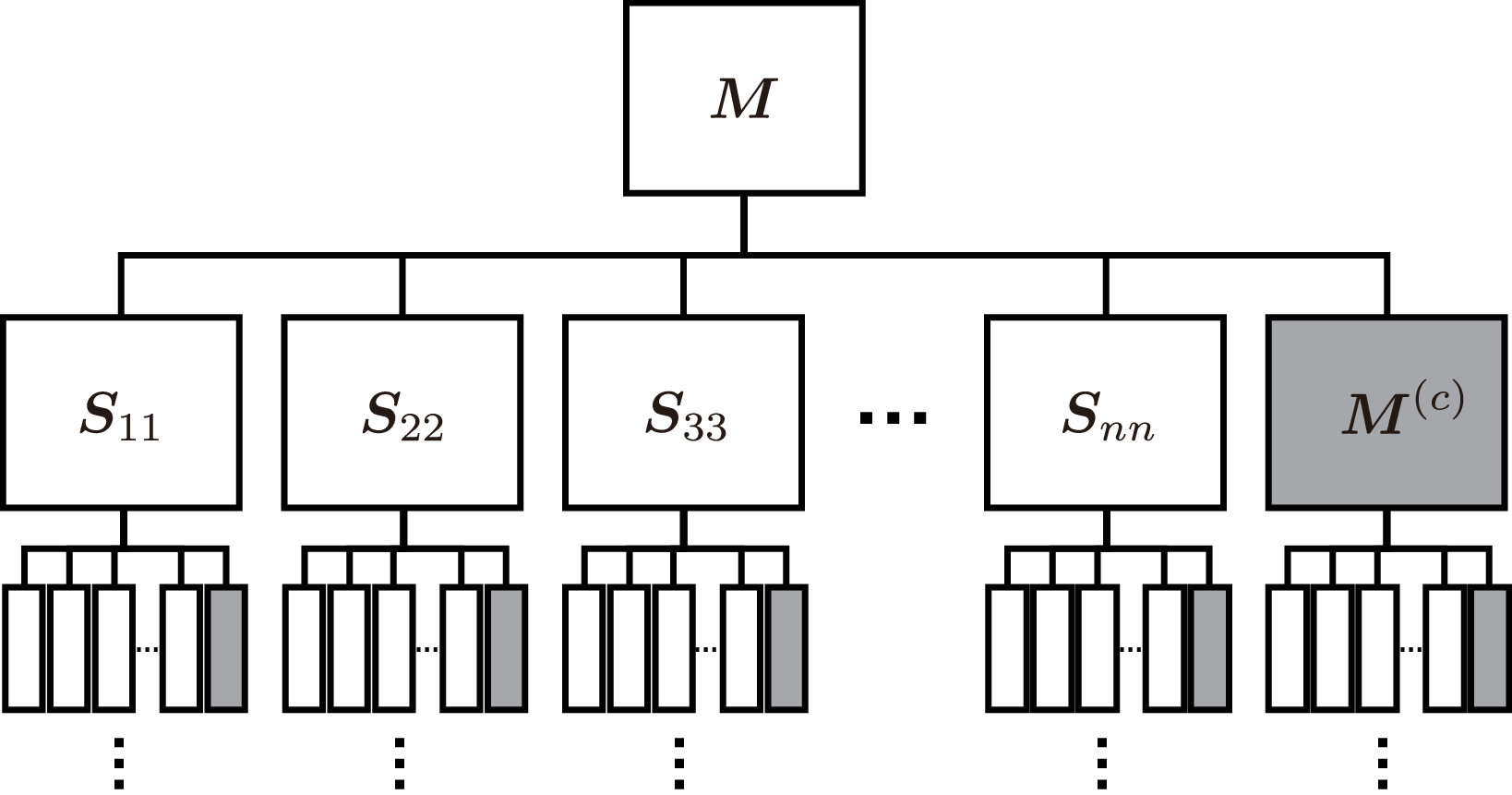}
\caption{
Tree structure of the submatrix partitioning used in the recursive construction of the basis set.
The complementary components of the eigenvectors associated with the principal submatrices are shown as shaded boxes.\\
}
\label{symfc2:Fig-block-tree}
\end{figure}

Based on the partitioning of orthogonal projection matrices and their products described in Sec.~\ref{symfc2:sec-projection-matrix-partitioning}, we develop a recursive algorithm for efficiently solving the associated eigenvalue problems.
In this approach, submatrix partitioning is recursively applied to previously partitioned submatrices, and eigenvalue problems for progressively smaller matrices are solved at each level.
Figure~\ref{symfc2:Fig-flowchart} shows the recursive algorithm for solving the eigenvalue problem of a given symmetric matrix $\bm{M}$.
The matrix $\bm{M}$ is assumed to be either an orthogonal projection matrix or a product of noncommuting orthogonal projection matrices.
The following steps are repeated recursively until the submatrices become sufficiently small to be solved directly using standard eigenvalue problem solvers.
At the same time, the submatrices must remain sufficiently large for the eigenvectors corresponding to the eigenvalue 1 of the principal submatrices to be identified.

(1) If the size of $\bm{M}$ exceeds the threshold value $\theta$, the procedure proceeds to step (2).
Otherwise, the eigenvalue problem of $\bm{M}$ is solved using a standard eigenvalue solver. 
As a result, the eigenvectors associated with the unit eigenvalue, denoted by $\bm{B_M}$, and the matrix 
\begin{equation}
\bm{M}^{(c)} = \bm{M} - \bm{B_M} \bm{B_M}^\top
\end{equation}
corresponding to the complementary vector space orthogonal to the space spanned by $\bm{B_M}$ are obtained.

(2) The matrix $\bm{M}$ is partitioned into $n \times n$ block submatrices.
The resulting block representation is given by
\begin{equation}
\label{symfc2:eq-submatrix-partitioning}
\bm{M} = 
\begin{bmatrix}
\bm{S}_{11} & \bm{S}_{12} & \cdots & \bm{S}_{1n} \\
\bm{S}_{21} & \bm{S}_{22} & \cdots & \bm{S}_{2n} \\
\vdots      & \vdots      & \ddots & \vdots      \\
\bm{S}_{n1} & \bm{S}_{n2} & \cdots & \bm{S}_{nn}
\end{bmatrix},
\end{equation}
where each principal submatrix $\bm{S}_{11}, \bm{S}_{22}, \ldots$ is a square matrix.
The eigenvalue problems associated with all principal submatrices are then solved recursively by returning to step (1).
Once these eigenvalue problems have been solved, the corresponding eigenvectors for the unit eigenvalue are obtained as
\begin{eqnarray}
\bm{S}_{11} \bm{B}_{11} &=& \bm{B}_{11} \nonumber \\
\bm{S}_{22} \bm{B}_{22} &=& \bm{B}_{22} \nonumber \\
& \vdots & \\
\bm{S}_{nn} \bm{B}_{nn} &=& \bm{B}_{nn}. \nonumber
\end{eqnarray}
Using the eigenvectors associated with the unit eigenvalue of the principal submatrices, the eigenvectors of the matrix $\bm{M}$ are constructed as
\begin{equation}
\bm{B_M}^{(1)} = 
\begin{bmatrix}
\bm{B}_{11} & \bm{0}      & \cdots & \bm{0}      \\
\bm{0}      & \bm{B}_{22} & \cdots & \bm{0}      \\
\vdots      & \vdots      & \ddots & \vdots      \\
\bm{0}      & \bm{0}      & \cdots & \bm{B}_{nn}
\end{bmatrix}
.
\end{equation}

(3) 
The complementary matrix 
\begin{equation}
\bm{M}^{(2)} = \bm{M} - \bm{B_M}^{(1)} \bm{B_M}^{(1)\top}
\end{equation}
is then calculated.
The eigenvalue problem for $\bm{M}^{(2)}$ is solved, and the eigenvectors corresponding to the unit eigenvalue are obtained as
\begin{equation}
\bm{M}^{(2)} \bm{B}_M^{(2)} = \bm{B}_M^{(2)}.
\end{equation}
In practice, an efficient procedure is employed to solve the eigenvalue problem for $\bm{M}^{(2)}$.
Details of this procedure are provided in Appendix~\ref{symfc2:appendix-complementary-projector}.

(4) 
The complete set of eigenvectors of $\bm{M}$ associated with the unit eigenvalue is given by
\begin{eqnarray}
\label{symfc2:eq-block-matrix-eigvecs}
\bm{B_M} & = & [\bm{B_M}^{(1)} \: \bm{B_M}^{(2)}] \nonumber \\
& = &
\begin{bmatrix}
\bm{B}_{11} & \bm{0}      & \cdots & \bm{0}      &  \\
\bm{0}      & \bm{B}_{22} & \cdots & \bm{0}      &  \\
\vdots      & \vdots      & \ddots & \vdots      & \bm{B_M}^{(2)} \\
\bm{0}      & \bm{0}      & \cdots & \bm{B}_{nn} &
\end{bmatrix}
.
\end{eqnarray}
In addition, the matrix corresponding to the complementary vector space orthogonal to the space spanned by $\bm{B}_M$ is calculated as
\begin{equation}
\bm{M}^{(c)} = \bm{M} - \bm{B}_M \bm{B}_M^\top.
\end{equation}

Figure~\ref{symfc2:Fig-block-tree} schematically illustrates the tree structure of the submatrix partitioning used in the recursive construction of the basis set.
Each partitioning step consists of solving the eigenvalue problems for the principal submatrices $\bm{S}_{ii}$, together with the eigenvalue problem for the complementary projection matrix $\bm{M}^{(2)}$.

Note that the proposed recursive approach has been implemented in the \textsc{symfc} code \cite{PhysRevB.110.214302,symfc-project}.
To reduce memory consumption and accelerate computations in both the construction of the force-constant basis set and the subsequent estimation of force constants, the implementation exploits the block-matrix structure of the basis set throughout the entire computational procedure.
Since most basis vectors are represented by small block matrices as shown in Eq. (\ref{symfc2:eq-block-matrix-eigvecs}), this formulation can significantly reduce memory requirements compared with standard solvers.
Moreover, although the procedures for solving eigenvalue problems of projection matrices are applied to force-constant basis-set construction in this study, the proposed approach is applicable to more general eigenvalue problems involving projection matrices.

\section{Results and discussion}
\label{symfc2:sec-results}

In this section, several applications of the proposed approach are demonstrated, including self-consistent phonon calculations for large low-symmetry systems, systematic basis-set construction for third-order force constants, lattice thermal conductivity calculations for low-symmetry compounds, and accurate estimation of force constants up to fourth order without introducing cutoff distances.
These applications involve computational tasks that are computationally demanding or even prohibitive with conventional methods, demonstrating the scalability and broad applicability of the improved projector-based framework.

\subsection{Self-consistent phonon calculations in low-symmetry large systems}

Effective second-order force constants play an essential role in self-consistent phonon calculations.
Because second-order supercell force constants contain only $9N^2$ elements, where $N$ denotes the number of atoms in the supercell, the number of independent force constants is generally small.
However, for large systems with only a small number of symmetry operations, the number of independent force constants can become very large even for second-order force constants.
As a result, constructing a complete force-constant basis set and determining force constants that satisfy all invariance conditions become computationally prohibitive using conventional approaches.
In this subsection, we present self-consistent phonon calculations for grain-boundary models as a representative application that requires large-scale basis-set construction and optimization of a large number of independent effective force constants.

We evaluate the temperature dependence of the excess free energies of two $\langle 110 \rangle$ symmetric tilt grain boundaries (STGBs) in elemental Al up to 1000 K.
The grain boundaries considered are the $\Sigma 3 \langle 110 \rangle$ STGBs with misorientation angles of 70.5$^\circ$ and 109.5$^\circ$.
Their calculation models are expressed by 96 and 192 atoms, respectively.
The optimized structures of these models were obtained by searching for the globally optimal microscopic structures with respect to rigid-body displacements for each set of macroscopic variables using a multistart optimization method together with a polynomial machine-learning potential (MLP) \cite{PhysRevMaterials.4.123607}.

We employ the stochastic self-consistent harmonic approximation (SSCHA) \cite{Errea-SSCHA-2013, van-Roekeghem-2020} to evaluate the grain-boundary excess free energy, including anharmonic vibrational contributions.
We evaluated the free energy values obtained from the SSCHA calculations using the definition of free energy adopted in Ref. \cite{Errea-SSCHA-2013}.
Thermal expansion effects are not taken into account in the present calculations.
An iterative SSCHA procedure \cite{van-Roekeghem-2020} implemented in the \textsc{pypolymlp} code \cite{doi:10.1063/5.0129045} is adopted to optimize the effective force constants, where harmonic contributions are evaluated using the \textsc{phonopy} code \cite{Togo20151,phonopy-phono3py-JPCM}.

SSCHA calculations were performed using $2 \times 2 \times 1$ supercell expansions for both STGB models.
As a result, the STGB supercell models with misorientation angles of 70.5$^\circ$ and 109.5$^\circ$ contain 384 and 768 atoms, respectively.
In addition, SSCHA calculations were also carried out using a $4 \times 4 \times 4$ expansion of the face-centered cubic (fcc) unit cell to evaluate the reference free energy required for calculating the excess free energy.
For the STGB model with a misorientation angle of 70.5$^\circ$, no cutoff distance was imposed on the force constants.
In contrast, a cutoff distance of 12~\AA{} was adopted for the STGB model with a misorientation angle of 109.5$^\circ$, because the longest axis of the corresponding supercell is approximately 80~\AA.

The numbers of basis vectors in the complete second-order force-constant basis sets are 7536 and 22935 for the STGB supercell models with misorientation angles of 70.5$^\circ$ and 109.5$^\circ$, respectively.
In each iteration of the SSCHA calculations, the numbers of structures sampled from the density matrix defined by the effective force constants at a given temperature were set to 20,000 and 40,000, respectively.
Using the forces and atomic displacements obtained from these structures, the expansion coefficients with respect to the basis sets were estimated using the least-squares technique implemented in the \textsc{symfc} code \cite{PhysRevB.110.214302,symfc-project}.
Once the expansion coefficients are obtained, the effective force constants can be calculated using Eq.~(\ref{symfc2:eq-fc-expansion}).

\begin{figure}[tbp]
\includegraphics[clip,width=\linewidth]{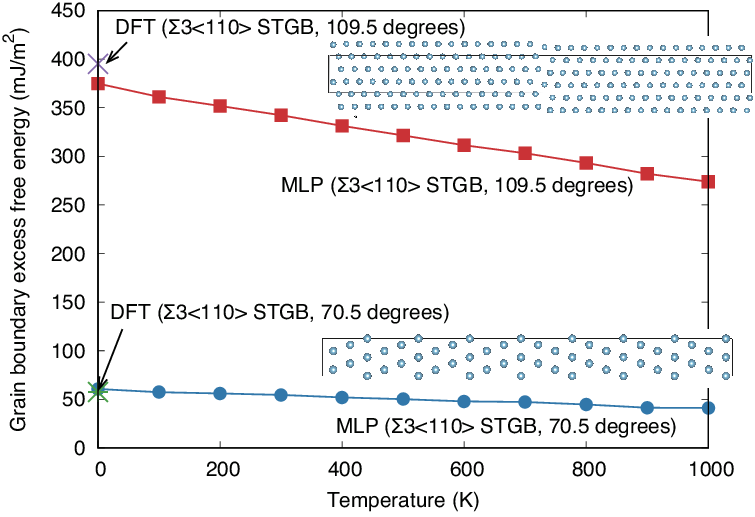}
\caption{
Temperature dependence of the grain-boundary excess free energy relative to that of the equilibrium fcc structure in elemental Al. 
The free energies were calculated using a combination of the polynomial MLP and the SSCHA approach. 
For comparison, grain-boundary excess energies calculated using DFT are also shown. 
These data are taken from Ref.~\cite{PhysRevMaterials.4.123607}.
}
\label{symfc2:Fig-free-energy-gb-Al}
\end{figure}

To evaluate the energies and forces for a large number of sampled structures in the SSCHA calculations, we employ a polynomial MLP \cite{PhysRevB.99.214108,doi:10.1063/5.0129045}.
The polynomial MLP represents the potential energy as a polynomial function of rotational polynomial invariants constructed from order parameters expressed in terms of radial functions and spherical harmonics.
In this study, we use a general-purpose polynomial MLP for elemental aluminum developed elsewhere \cite{doi:10.1063/5.0129045} and distributed through the polynomial MLP repository \cite{MachineLearningPotentialRepository}.
Although this MLP has been updated from the version used for calculating grain-boundary excess energies in Ref.~\cite{PhysRevMaterials.4.123607} by incorporating a wider variety of structures into the training dataset, the current and previous MLPs yield nearly identical grain-boundary excess energies and atomic structures for the grain-boundary models.

Figure~\ref{symfc2:Fig-free-energy-gb-Al} shows the grain-boundary excess free energy relative to the free energy of the equilibrium fcc structure of elemental Al.
Since the effective force constants contain a large number of independent components and are estimated from extensive force--displacement datasets, the computational cost of force-constant estimation remains high throughout the SSCHA calculations.
Nevertheless, the calculated excess free energies exhibit a smooth temperature dependence.
These results demonstrate that the present procedure enables SSCHA calculations for large systems with extensive force-constant basis sets, allowing the evaluation of vibrational properties including anharmonic contributions.

Regarding the computational cost, it is worth noting that the computational cost of energy and force evaluations using the polynomial MLP is negligible in each SSCHA iteration. When an extensive dataset is used to estimate the effective force constants, as in the present study, where the datasets consist of 20,000 and 40,000 structures, the main computational bottleneck is the estimation of the effective force constants from the force–displacement datasets.

\subsection{Third-order force-constant basis sets}
\label{symfc2:sec-distribution-fc3}

Constructing complete basis sets for third-order force constants using conventional approaches is computationally prohibitive for many compounds.
To demonstrate the broad practical applicability of the proposed procedure, we constructed complete basis sets for the third-order supercell force constants of compounds included in \textsc{Phonondb} \cite{Togo2023MDRPhonon}. 
In total, basis sets were generated for 9,587 compounds. 
Nearly isotropic supercells were constructed from the unit cells of these compounds, with their sizes chosen such that the number of atoms ranged from 32 to 128, except for compounds whose unit cells already contained more than 128 atoms.
No cutoff distance was imposed on the force constants.

\begin{figure}[tbp]
\includegraphics[clip,width=\linewidth]{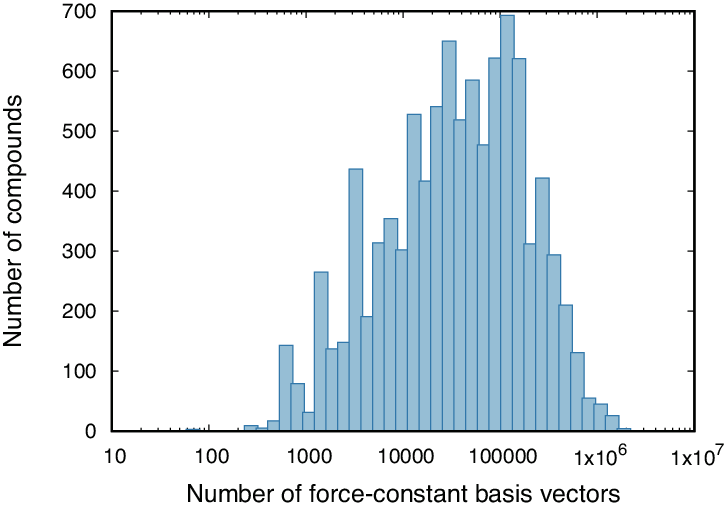}
\caption{
Number of compounds in \textsc{Phonondb} \cite{Togo2023MDRPhonon} as a function of the number of basis vectors for the third-order force constants.
The numbers of basis vectors were calculated using the procedure proposed in this study. 
These basis vectors constitute complete basis sets that satisfy all invariance conditions.
}
\label{symfc2:Fig-basis-histogram-fc3}
\end{figure}

Figure \ref{symfc2:Fig-basis-histogram-fc3} shows the distribution of compounds as a function of the number of force-constant basis vectors. 
The computational times required to construct complete basis sets for third-order force constants on a single standard workstation are summarized in Appendix \ref{symfc2:sec-appendix-time-fc3}.
The size of the third-order force-constant basis sets ranges from approximately $10^2$ to $10^6$.
Approximately 78\% of the compounds have basis sets containing more than $10^4$ vectors, making them computationally challenging to obtain using conventional approaches.

Note that the complete basis sets for the third-order force constants in these compounds are considerably larger than those typically encountered in simple crystal structures such as rocksalt, zincblende, and wurtzite systems.
For example, the rocksalt structure with a $2\times2\times2$ supercell containing 64 atoms requires only 758 basis functions for the third-order force constants, whereas the corresponding number for the zincblende structure is 1536.
Even in the wurtzite structure with a relatively large $3\times3\times2$ supercell containing 72 atoms, the basis size remains moderate at 7752.
In such highly symmetric systems, both basis-set construction and force-constant estimation can be carried out efficiently using the previous projector-based procedure together with a standard eigenvalue solver.
By contrast, as illustrated in Fig.~\ref{symfc2:Fig-basis-histogram-fc3}, many compounds exhibit substantially larger basis sets, making the present procedure particularly effective for systematically imposing all invariant conditions on the third-order force constants.

\subsection{Lattice thermal conductivity in low-symmetry compounds}

We next demonstrate LTC calculations for low-symmetry compounds with only a small number of space-group operations, which results in a large number of independent third-order force constants.
As representative examples, we consider five compounds---Cu$_8$SiS$_6$, KGePO$_5$, CaSiO$_3$, Al(OH)$_3$, and Li$_6$Ge$_2$O$_7$---and perform LTC calculations for these systems.
Their space groups are $Pmn2_1$, $Pna2_1$, $C2/c$, $P2_1/c$, and $P2_1/c$, respectively. 
The numbers of atoms in the unit cells are 30, 64, 60, 56, and 60, respectively. 
Supercells were constructed using $2\times2\times1$, $1\times2\times1$, $1\times2\times1$, $1\times2\times1$, and $1\times2\times1$ expansions, respectively, to evaluate supercell force constants.
In addition, no cutoff distance was introduced for either the second- or third-order force constants. 
Consequently, the sizes of the third-order force-constant basis sets are 478876, 1161232, 477522, 776258, and 955905, respectively.

To develop a force--displacement dataset for estimating second- and third-order force constants, we construct an on-the-fly polynomial MLP following the procedure described in Ref.~\cite{10.1063/5.0211296}.
First, a DFT-based force--displacement dataset was generated using DFT calculations solely for the development of the on-the-fly polynomial MLP.
The dataset consists of 200 supercell structures with atomic displacements of 0.03~\AA\ for the supercell sizes described above. 
DFT calculations were performed using the plane-wave projector augmented-wave (PAW) method \cite{PAW1} within the Perdew--Burke--Ernzerhof (PBE) exchange-correlation functional \cite{GGA:PBE96}, as implemented in the \textsc{vasp} code \cite{VASP1,VASP2,PAW2}.
Based on the DFT dataset, a polynomial MLP was developed using linear ridge regression implemented in the \textsc{pypolymlp} code.

The developed polynomial MLP was then used to efficiently evaluate atomic forces for supercell structures of each compound, thereby generating an MLP-based force--displacement dataset that was substantially larger than the original DFT dataset.
We employed the brute-force finite-displacement method implemented in \textsc{phono3py} \cite{phono3py,phonopy-phono3py-JPCM}.
The equilibrium atomic positions were optimized using the polynomial MLP so that the residual forces at the equilibrium structures were negligibly small.
Supercell structures with atomic displacements were then generated from the optimized equilibrium structures, and the force constants were estimated from the resulting force--displacement dataset.
The projection matrices constructed using the present procedure were subsequently applied to the estimated force constants using Eq.~(\ref{symfc2:eq-projection-apply}) to enforce all symmetry and invariance conditions.

\begin{figure}[tbp]
\includegraphics[clip,width=\linewidth]{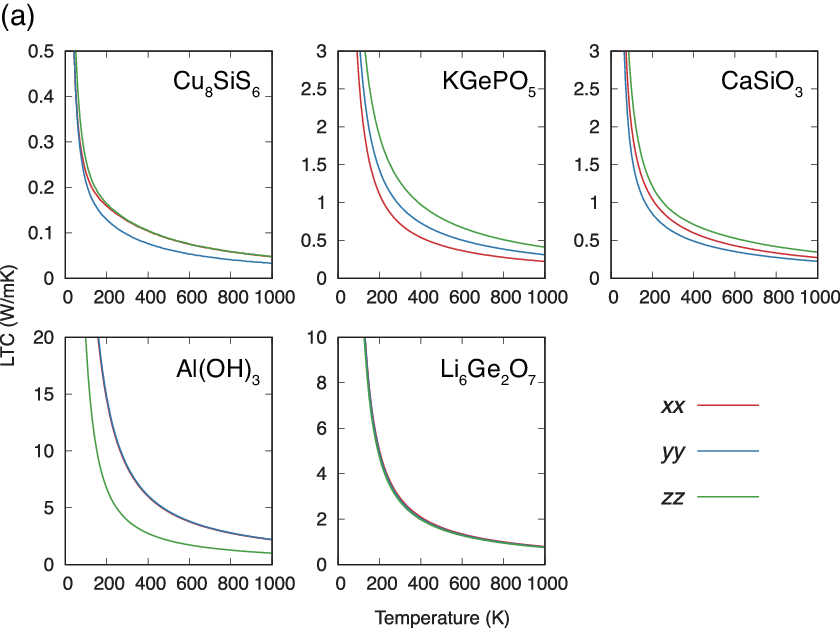}
\includegraphics[clip,width=\linewidth]{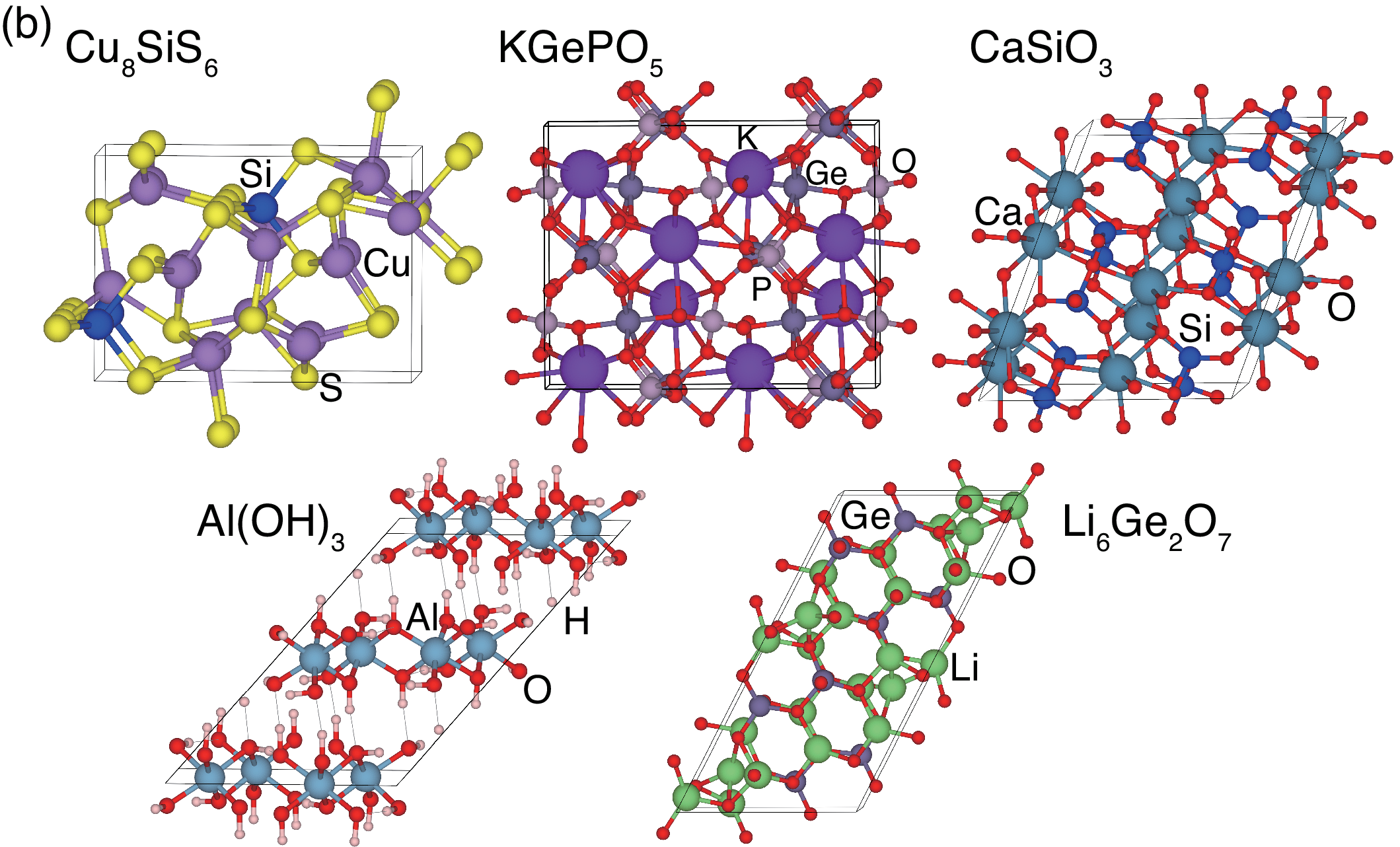}
\caption{
(a) Temperature dependence of the LTC for five compounds with a large number of independent third-order force constants.
The force constants are estimated from force–displacement datasets constructed using on-the-fly polynomial MLPs.
Only diagonal elements are shown.
(b) Crystal structures of the unit cells of the five compounds.
}
\label{symfc2:Fig-ltc-panels}
\end{figure}

Figure \ref{symfc2:Fig-ltc-panels} shows the temperature dependence of the LTC for the five compounds, calculated using the symmetrized force constants.
The LTCs were calculated by solving the Peierls--Boltzmann transport equation within the relaxation time approximation \cite{Peierls-Boltzmann-1929,Peierls-Quantum-Theory-of-Solids,Allen-LTC-2018} using the \textsc{phono3py} code \cite{phono3py,phonopy-phono3py-JPCM}.
The LTC calculations were simplified by considering only phonon--phonon scattering in the evaluation of the phonon relaxation times. 
The phonon lifetimes were computed from the supercell force constants as the reciprocals of the imaginary part of the phonon self-energy associated with the bubble diagram, and these values were used as the relaxation times.
Phonon properties required for the LTC calculations were obtained from the second-order supercell force constants. 
Even for compounds with a large number of force constants, third-order force constants satisfying invariant conditions can be calculated efficiently using the proposed procedure.
As a consequence, LTC calculations using the complete force-constant basis set can be applied to compounds with complex structures and compositions.

\subsection{Fourth-order force constants}

\begin{figure}[tbp]
\includegraphics[clip,width=\linewidth]{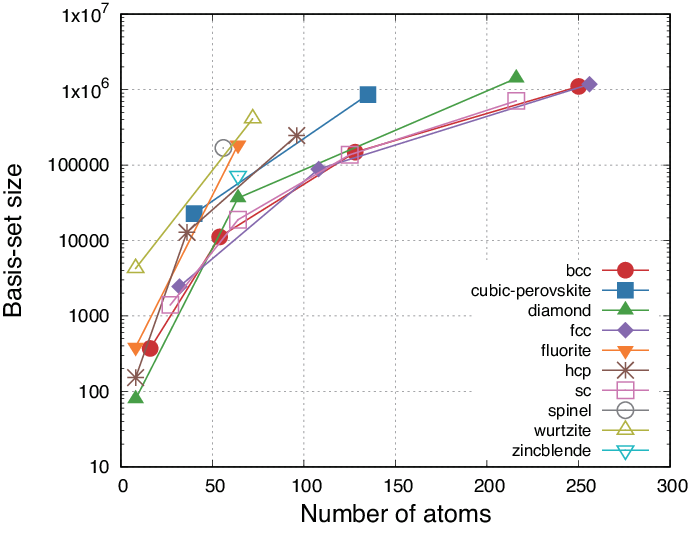}
\caption{
Sizes of complete fourth-order force-constant basis sets for common crystal structures, including body-centered cubic (bcc), cubic perovskite, diamond, face-centered cubic (fcc), fluorite, hexagonal close-packed (hcp), simple cubic (sc), spinel, wurtzite, and zincblende.
Their supercell dependencies are shown. 
}
\label{symfc2:Fig-basis-size-fc4}
\end{figure}

Constructing complete basis sets for fourth-order force constants can also be performed efficiently using the present procedure.
We constructed complete basis sets for fourth-order force constants for common elemental and ionic crystal structures.
No cutoff radii were introduced in these calculations.
Several supercell sizes were considered to illustrate the dependence of the basis-set size on the number of atoms in the supercell.

Figure~\ref{symfc2:Fig-basis-size-fc4} shows the number of fourth-order force-constant basis vectors in the complete basis sets for these crystal structures.
For fourth-order force constants, even in compounds with many space-group operations, the size of the complete basis set becomes large and exceeds 30000 for standard supercell sizes unless a cutoff distance is introduced.
Nevertheless, complete basis sets for fourth-order force constants can be constructed with reasonable computational resources.

We then demonstrate the estimation of fourth-order supercell force constants together with second- and third-order force constants.
The force constants are computed for diamond silicon using a $2\times2\times2$ supercell containing 64 atoms. 
No cutoff radius is applied to the second-, third-, or fourth-order force constants.
Consequently, the complete force-constant basis set contains 37,326 independent basis functions, corresponding to the cumulative number of basis functions up to fourth order.

To construct the force--displacement datasets used for estimating the force constants, we employ polynomial MLPs. 
For elemental silicon, we use a polynomial MLP developed in Ref.~\cite{FUJII2022111137} and available in the repository \cite{MachineLearningPotentialRepository}, which has been shown to achieve high predictive accuracy for LTC calculations.
Force--displacement datasets containing 500--20,000 supercell structures are generated from the equilibrium structure by applying random atomic displacements with amplitudes of 0.01, 0.03, 0.05, 0.1, 0.2, 0.3, 0.4, and 0.5~\AA\ to all atoms. 
Each combination of dataset size and displacement amplitude is considered independently.
The atomic forces for these structures are then efficiently evaluated using the polynomial MLP.
Finally, the resulting force--displacement datasets are used to simultaneously estimate the independent coefficients associated with the complete basis sets for the second-, third-, and fourth-order force constants by means of the least-squares method implemented in the \textsc{symfc} code.

\begin{figure}[tbp]
\includegraphics[clip,width=\linewidth]{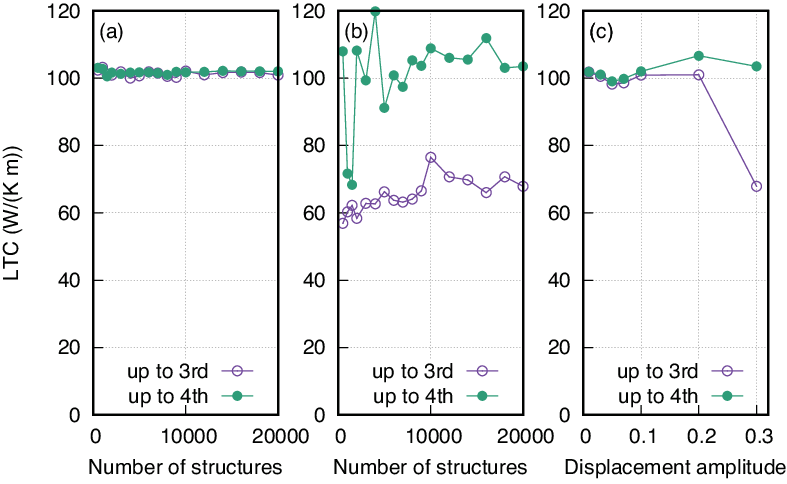}
\caption{
Dependence of the LTC at 300 K on the number of supercell structures included in the force–displacement dataset for diamond silicon, using atomic displacement amplitudes of (a) 0.1 and (b) 0.3 \AA. 
The LTC values calculated using force constants estimated up to third order are compared with those obtained using force constants estimated up to fourth order.
(c) Dependence of the LTC at 300 K on the atomic displacement amplitude. 
A total of 20,000 supercell structures is used in this panel.
}
\label{symfc2:Fig-ltc-4th-Si}
\end{figure}

Figures \ref{symfc2:Fig-ltc-4th-Si}(a) and (b) show the dependence of the LTC at 300 K on the number of supercell structures included in the force--displacement dataset. 
In panels (a) and (b), the atomic displacement amplitudes are fixed at 0.1 and 0.3~\AA, respectively. 
The LTC values are evaluated both with and without fourth-order force constants for comparison.

When a displacement amplitude of 0.1~\AA\ is used, the LTC values converge with respect to the number of supercell structures even when only a small number of structures is included in the dataset.
As the displacement amplitude increases, higher-order anharmonic contributions are effectively renormalized into the second- and third-order force constants when fourth-order force constants are not included.
As a result, the calculated LTC values deviate significantly from those obtained when fourth-order force constants are explicitly taken into account.

In contrast, when fourth-order force constants are included in the force constant estimation, the calculated LTC values are close to those obtained from a force--displacement dataset with a smaller displacement amplitude.
This indicates that accurate estimation of the fourth-order force constants enables the second- and third-order force constants to be extracted accurately even from force--displacement datasets with relatively large atomic displacements.
Nevertheless, a larger number of supercell structures is required to obtain converged force constants and LTC values.

Figure \ref{symfc2:Fig-ltc-4th-Si}(c) shows the dependence of the LTC on the amplitude of the atomic displacements. 
The number of supercell structures is fixed at 20,000.
For datasets with small displacement amplitudes, the calculated LTC is approximately 100--105~W/Km, regardless of whether fourth-order force constants are included.
In contrast, for datasets with larger displacement amplitudes, the calculated LTC depends strongly on whether fourth-order force constants are included.
When fourth-order force constants are estimated together with the second- and third-order force constants, the calculated LTC remains close to the values obtained from datasets with small displacement amplitudes. 
This result indicates that the second-, third-, and fourth-order force constants can all be accurately determined through the simultaneous estimation up to fourth order.

\section{Conclusion}
\label{symfc2:sec-conclusion}

In this work, we developed an efficient eigenvalue solver for projection matrices and incorporated it into the projector-based framework for force-constant construction.
Specifically, the proposed algorithm is based on partitioning a projection matrix into submatrices and solving the eigenvalue problems associated with these smaller submatrices.
Furthermore, we introduced a recursive algorithm in which the matrix-partitioning procedure is applied repeatedly, thereby exploiting the block structure of the basis set throughout the entire computational procedure to further reduce the required computational resources.
This advancement substantially reduces the computational cost and significantly broadens the applicability of the method.
In particular, it enables the practical treatment of large-scale systems, low-symmetry crystals, and higher-order force constants that are otherwise difficult to handle using conventional approaches.

Several applications of the proposed approach have been demonstrated, including self-consistent phonon calculations for large low-symmetry systems, systematic basis-set construction for third-order force constants, LTC calculations for low-symmetry compounds, and accurate estimation of force constants up to fourth order without introducing cutoff distances.
These applications, which are computationally demanding or prohibitive for conventional methods, demonstrate the capability of the improved projector-based framework to perform efficient and robust force-constant calculations for large and complex materials systems.

\begin{acknowledgments}
This work was supported by JSPS KAKENHI Grant Numbers 26K01193, 25K22169, 25H00420, JP24K08021, JP24H00190, JP25H01246, and 25H01252.
\end{acknowledgments}

\appendix

\section{Efficient algorithm for solving complementary projection matrix}
\label{symfc2:appendix-complementary-projector}

As described in Secs.~\ref{symfc2:sec-projection-matrix-partitioning} and \ref{symfc2:sec-recursive-algorithm}, after solving the eigenvalue problems for the principal submatrices and obtaining the eigenvectors corresponding to the unit eigenvalue, denoted by $\bm{B}^{(1)}$ or $\bm{B}_{\bm{M}}^{(1)}$, the remaining eigenvectors associated with the unit eigenvalue must also be determined to complete the basis for the corresponding eigenspace.
Following the notation introduced in Sec.~\ref{symfc2:sec-recursive-algorithm}, the eigenvalue problem for
\begin{equation}
\bm{M}^{(2)} = \bm{M} - \bm{B}_{\bm{M}}^{(1)} \bm{B}_{\bm{M}}^{(1)\top}
\end{equation}
must be solved, and the eigenvectors corresponding to the unit eigenvalue, $\bm{M}^{(2)} \bm{B}_{\bm{M}}^{(2)} = \bm{B}_{\bm{M}}^{(2)}$, are obtained.

In this appendix, we present an efficient procedure for solving the eigenvalue problem for $\bm{M}^{(2)}$ and constructing a reduced matrix of $\bm{M}^{(2)}$ by utilizing the eigenvectors corresponding to the nonnegative eigenvalues less than one obtained from the eigenvalue problems for the principal submatrices.
When solving the eigenvalue problems for the principal submatrices $\bm{S}_{11}, \bm{S}_{22}, \ldots, \bm{S}_{nn}$, both the eigenvectors corresponding to the unit eigenvalue, $\bm{B}_{11}, \bm{B}_{22}, \ldots, \bm{B}_{nn}$, and those corresponding to non-negative eigenvalues smaller than one, $\bm{C}_{11}, \bm{C}_{22}, \ldots, \bm{C}_{nn}$, are obtained.
Accordingly, their eigendecompositions are written as
\begin{eqnarray}
\label{symfc2:eq-eigendecompositions-submatrices}
\bm{S}_{11} &=& \bm{B}_{11} \bm{B}_{11}^\top + \bm{C}_{11} \bm{\Lambda}_{11} \bm{C}_{11}^\top \nonumber \\
\bm{S}_{22} &=& \bm{B}_{22} \bm{B}_{22}^\top + \bm{C}_{22} \bm{\Lambda}_{22} \bm{C}_{22}^\top \nonumber \\
& \vdots & \\
\bm{S}_{nn} &=& \bm{B}_{nn} \bm{B}_{nn}^\top + \bm{C}_{nn} \bm{\Lambda}_{nn} \bm{C}_{nn}^\top, \nonumber
\end{eqnarray}
where the notation is consistent with that used in Eq.~(\ref{symfc2:eq-eigendecomposition}).
When the eigenvectors $\bm{B}_{11}, \bm{B}_{22}, \ldots, \bm{B}_{nn}$ and $\bm{C}_{11}, \bm{C}_{22}, \ldots, \bm{C}_{nn}$ are assembled into the matrices $\bm{B}_{\bm{M}}^{(1)}$ and $\bm{C}_{\bm{M}}^{(1)}$, respectively, as follows,
\begin{equation}
\bm{B}_{\bm{M}}^{(1)} = 
\begin{bmatrix}
\bm{B}_{11} & \bm{0}      & \cdots & \bm{0}      \\
\bm{0}      & \bm{B}_{22} & \cdots & \bm{0}      \\
\vdots      & \vdots      & \ddots & \vdots      \\
\bm{0}      & \bm{0}      & \cdots & \bm{B}_{nn}
\end{bmatrix}
\end{equation}
and 
\begin{equation}
\bm{C}_{\bm{M}}^{(1)} = 
\begin{bmatrix}
\bm{C}_{11} & \bm{0}      & \cdots & \bm{0}      \\
\bm{0}      & \bm{C}_{22} & \cdots & \bm{0}      \\
\vdots      & \vdots      & \ddots & \vdots      \\
\bm{0}      & \bm{0}      & \cdots & \bm{C}_{nn}
\end{bmatrix}
,
\end{equation}
the matrix $\bm{M}^{(2)}$ is represented as
\begin{widetext}
\begin{eqnarray}
\bm{M}^{(2)} 
&=& \bm{M} - \bm{B}_{\bm{M}}^{(1)} \bm{B}_{\bm{M}}^{(1)\top} \nonumber \\
&=& 
\left( \bm{B}_{\bm{M}}^{(1)} \bm{B}_{\bm{M}}^{(1)\top} + \bm{C}_{\bm{M}}^{(1)} \bm{C}_{\bm{M}}^{(1)\top} \right) 
\left( \bm{M} - \bm{B}_{\bm{M}}^{(1)} \bm{B}_{\bm{M}}^{(1)\top} \right) 
\left( \bm{B}_{\bm{M}}^{(1)} \bm{B}_{\bm{M}}^{(1)\top} + \bm{C}_{\bm{M}}^{(1)} \bm{C}_{\bm{M}}^{(1)\top} \right) 
\\
&=& \bm{C}_{\bm{M}}^{(1)} \left[ \bm{C}_{\bm{M}}^{(1)\top} \bm{M} \bm{C}_{\bm{M}}^{(1)}\right] \bm{C}_{\bm{M}}^{(1)\top}. \nonumber
\end{eqnarray}
\end{widetext}
Here, the orthonormality relations
\begin{equation}
\bm{C}_{\bm{M}}^{(1)\top} \bm{C}_{\bm{M}}^{(1)} = \bm{I}_{\bm{C}_{\bm{M}}^{(1)}},\:\: \bm{C}_{\bm{M}}^{(1)\top} \bm{B}_{\bm{M}}^{(1)} = \bm{0}
\end{equation}
together with the identity
\begin{equation}
\bm{I} = \bm{B}_{\bm{M}}^{(1)} \bm{B}_{\bm{M}}^{(1)\top} + \bm{C}_{\bm{M}}^{(1)} \bm{C}_{\bm{M}}^{(1)\top}
\end{equation}
are used, in addition to the relation $\bm{M} \bm{B}_{\bm{M}}^{(1)} = \bm{B}_{\bm{M}}^{(1)}$, to derive the final equality.
The matrix $\bm{I}_{\bm{C}_{\bm{M}}^{(1)}}$ is the identity matrix of the same column dimension as $\bm{C}_{\bm{M}}^{(1)}$.

Since the size of $\bm{C}_{\bm{M}}^{(1)\top} \bm{M} \bm{C}_{\bm{M}}^{(1)}$ is generally much smaller than those of $\bm{M}$ and $\bm{M}^{(2)}$, it can be regarded as a reduced representation of $\bm{M}^{(2)}$.
Using the eigendecompositions of the principal submatrices, the matrix $\bm{C}_{\bm{M}}^{(1)\top} \bm{M} \bm{C}_{\bm{M}}^{(1)}$ can be computed as
\begin{eqnarray}
& & \bm{C}_{\bm{M}}^{(1)\top} \bm{M} \bm{C}_{\bm{M}}^{(1)} \nonumber\\
&=& 
\begin{bmatrix}
\bm{\Lambda}_{11} & \bm{C}_{11}^\top \bm{S}_{12} \bm{C}_{22} & \cdots & \bm{C}_{11}^\top \bm{S}_{1n} \bm{C}_{nn}      \\
\bm{C}_{22}^\top \bm{S}_{21} \bm{C}_{11} & \bm{\Lambda}_{22} & \cdots & \bm{C}_{22}^\top \bm{S}_{2n} \bm{C}_{nn}      \\
\vdots      & \vdots      & \ddots & \vdots      \\
\bm{C}_{nn}^\top \bm{S}_{n1} \bm{C}_{11}  & \bm{C}_{nn}^\top \bm{S}_{n2} \bm{C}_{22} & \cdots & \bm{\Lambda}_{nn}
\end{bmatrix}
.
\end{eqnarray}
Once the reduced matrix has been constructed, its eigenvalue problem is solved recursively using the same procedure shown in Sec. \ref{symfc2:sec-recursive-algorithm}.
Consequently, the eigenvectors corresponding to the unit eigenvalue of the compressed matrix are obtained as
\begin{equation}
\left[
\bm{C}_{\bm{M}}^{(1)\top}
\bm{M}
\bm{C}_{\bm{M}}^{(1)}
\right]
\bm{B}'
=
\bm{B}' .
\end{equation}
The eigenvectors of $\bm{M}^{(2)}$ corresponding to the unit eigenvalue are then recovered from $\bm{B}'$ as
\begin{equation}
\bm{B}_{\bm{M}}^{(2)}
=
\bm{C}_{\bm{M}}^{(1)} \bm{B}' .
\end{equation}
The complete basis set for $\bm{M}$ corresponding to the unit eigenvalue is thus obtained as the column-stacked matrix
\[
\bm{B}_{\bm{M}} = [\bm{B}_{\bm{M}}^{(1)} : \bm{B}_{\bm{M}}^{(2)}].
\]

Furthermore, the eigenvectors of $\bm{M}$ corresponding to non-negative eigenvalues smaller than one are identical to those of $\bm{M}^{(2)}$ corresponding to non-negative eigenvalues smaller than one.
Therefore, they can be obtained by solving the eigenvalue problem for $\bm{M}^{(2)}$.
The eigenvectors corresponding to the non-negative eigenvalues smaller than one for the reduced matrix are obtained from
\begin{equation}
\left[
\bm{C}_{\bm{M}}^{(1)\top}
\bm{M}
\bm{C}_{\bm{M}}^{(1)}
\right]
\bm{C}'
=
\bm{C}' \bm{\Lambda}_c',
\end{equation}
where $\bm{\Lambda}_c'$ is the diagonal matrix whose diagonal elements are the eigenvalues associated with $\bm{C}'$.
The eigenvectors of $\bm{M}$ corresponding to the non-negative eigenvalues smaller than one are then given by
\begin{equation}
\bm{C}_{\bm{M}}
=
\bm{C}_{\bm{M}}^{(1)} \bm{C}'.
\end{equation}

\section{Proof for product of noncommutative projection matrices}
\label{symfc2:appendix-proof2}

Let us consider the product of noncommutative orthogonal projection matrices $\bm{P}_A$ and $\bm{P}_B$, defined as $\bm{M} = \bm{P}_A \bm{P}_B \bm{P}_A$.
In this appendix, we show that, when the matrix $\bm{M}$ is partitioned into submatrices as in Eq.~(\ref{symfc2:eq-submatrix-partitioning}), the eigenvectors of a principal submatrix associated with the unity eigenvalue also constitute eigenvectors of both $\bm{M}$ and 
$\bm{P}_{A \cap B} = \lim_{p \to \infty} \bm{M}^p$
with the unit eigenvalue.

As shown in Sec.~\ref{symfc2:sec-projection-matrix-partitioning}, a principal submatrix can be expressed as $\bm{S}_{11} = \bm{P}_1 \bm {M} \bm{P}_1$, where $\bm{P}_1$ is the projection matrix used to extract the submatrix $\bm{S}_{11}$ from $\bm{M}$ defined in Eq. (\ref{symfc2:eq-p1matrix}). 
The projection matrix onto the common subspace simultaneously projected by
$\bm{P}_1$, $\bm{P}_A$, and $\bm{P}_B$ is given by \cite{pyle1967generalized}
\begin{equation}
\bm{P}_{A \cap B \cap 1} = \lim_{p \to \infty} \left( \bm{P}_1 \bm{M} \bm{P}_1 \right)^p.
\end{equation}
On the other hand, the same projection matrix can also be expressed as
\begin{equation}
\bm{P}_{A \cap B \cap 1} = \lim_{q \to \infty} \left( \bm{P}_1 \bm{P}_{A \cap B} \bm{P}_1 \right)^q.
\end{equation}
Therefore, the eigenvectors of $\bm{P}_1 \bm{M} \bm{P}_1$ associated with the unit eigenvalue also correspond to eigenvectors of
$\bm{P}_1 \bm{P}_{A \cap B} \bm{P}_1$
associated with the unit eigenvalue.

The eigenvectors of $\bm{P}_1 \bm{P}_{A \cap B} \bm{P}_1$ associated with the unit eigenvalue are also eigenvectors of $\bm{P}_{A \cap B}$ (and $\bm{P}_1$) associated with the unit eigenvalue.
Since $\bm{P}_{A \cap B} = \lim_{p \to \infty} \bm{M}^p$, the eigenvectors of $\bm{P}_{A \cap B}$ associated with the unit eigenvalue are also eigenvectors of $\bm{M}$.
Consequently, even when $\bm{M}$ is given by a product of noncommutative orthogonal projection matrices, the eigenvectors of a principal submatrix associated with the unit eigenvalue can be used to construct eigenvectors of $\bm{M}$ associated with the unit eigenvalue.

\section{Computational performance of basis-set construction}
\label{symfc2:sec-appendix-time-fc3}

\begin{figure}[tbp]
\includegraphics[clip,width=0.9\linewidth]{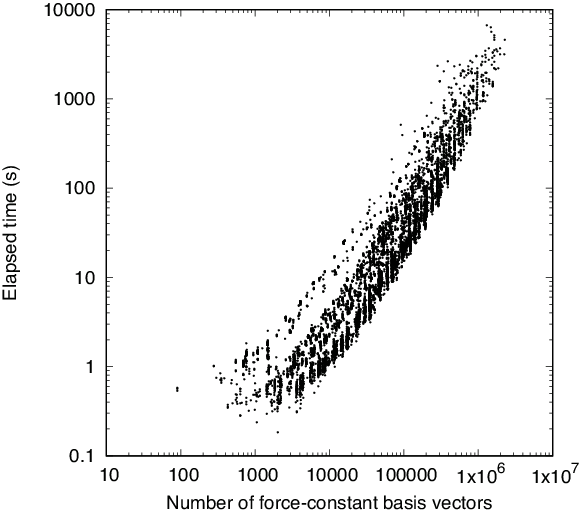}
\caption{
Distribution of computational times required to construct complete basis sets for the third-order force constants of 9587 compounds.
The computational times were estimated using a workstation equipped with two Intel(R) Xeon(R) Silver 4514Y CPUs and 128~GB of memory.
}
\label{symfc2:Fig-n-basis-time-fc3}
\end{figure}

Figure \ref{symfc2:Fig-n-basis-time-fc3} shows the estimated computational times required to construct complete basis sets for the third-order force constants of 9587 compounds.
The computational time exhibits a strong correlation with the size of the force-constant basis set. 
The number of basis vectors ranges from approximately $10^2$ to $2\times10^6$, while the construction of a complete basis set can be completed within one hour for a single compound.
For compounds with large basis sets, the present procedure substantially reduces the computational cost of constructing complete basis sets.

\bibliography{symfc}

\begin{thebibliography}{41}%
\makeatletter
\providecommand \@ifxundefined [1]{%
 \@ifx{#1\undefined}
}%
\providecommand \@ifnum [1]{%
 \ifnum #1\expandafter \@firstoftwo
 \else \expandafter \@secondoftwo
 \fi
}%
\providecommand \@ifx [1]{%
 \ifx #1\expandafter \@firstoftwo
 \else \expandafter \@secondoftwo
 \fi
}%
\providecommand \natexlab [1]{#1}%
\providecommand \enquote  [1]{``#1''}%
\providecommand \bibnamefont  [1]{#1}%
\providecommand \bibfnamefont [1]{#1}%
\providecommand \citenamefont [1]{#1}%
\providecommand \href@noop [0]{\@secondoftwo}%
\providecommand \href [0]{\begingroup \@sanitize@url \@href}%
\providecommand \@href[1]{\@@startlink{#1}\@@href}%
\providecommand \@@href[1]{\endgroup#1\@@endlink}%
\providecommand \@sanitize@url [0]{\catcode `\\12\catcode `\$12\catcode
  `\&12\catcode `\#12\catcode `\^12\catcode `\_12\catcode `\%12\relax}%
\providecommand \@@startlink[1]{}%
\providecommand \@@endlink[0]{}%
\providecommand \url  [0]{\begingroup\@sanitize@url \@url }%
\providecommand \@url [1]{\endgroup\@href {#1}{\urlprefix }}%
\providecommand \urlprefix  [0]{URL }%
\providecommand \Eprint [0]{\href }%
\providecommand \doibase [0]{http://dx.doi.org/}%
\providecommand \selectlanguage [0]{\@gobble}%
\providecommand \bibinfo  [0]{\@secondoftwo}%
\providecommand \bibfield  [0]{\@secondoftwo}%
\providecommand \translation [1]{[#1]}%
\providecommand \BibitemOpen [0]{}%
\providecommand \bibitemStop [0]{}%
\providecommand \bibitemNoStop [0]{.\EOS\space}%
\providecommand \EOS [0]{\spacefactor3000\relax}%
\providecommand \BibitemShut  [1]{\csname bibitem#1\endcsname}%
\let\auto@bib@innerbib\@empty
\bibitem [{\citenamefont {Dove}(1993)}]{IntroductionToLatticeDynamics}%
  \BibitemOpen
  \bibfield  {author} {\bibinfo {author} {\bibfnamefont {M.~T.}\ \bibnamefont
  {Dove}},\ }\href@noop {} {\emph {\bibinfo {title} {Introduction to Lattice
  Dynamics}}}\ (\bibinfo  {publisher} {Cambridge University Press},\ \bibinfo
  {year} {1993})\BibitemShut {NoStop}%
\bibitem [{\citenamefont {Wallace}(1998)}]{Thermodynamics-of-crystals}%
  \BibitemOpen
  \bibfield  {author} {\bibinfo {author} {\bibfnamefont {Duane~C.}\
  \bibnamefont {Wallace}},\ }\href@noop {} {\emph {\bibinfo {title}
  {Thermodynamics of crystals}}}\ (\bibinfo  {publisher} {Dover Publications},\
  \bibinfo {year} {1998})\BibitemShut {NoStop}%
\bibitem [{\citenamefont {Ziman}(1960)}]{Ziman-electrons-phonons}%
  \BibitemOpen
  \bibfield  {author} {\bibinfo {author} {\bibfnamefont {John~Michael}\
  \bibnamefont {Ziman}},\ }\href {\doibase
  10.1093/acprof:oso/9780198507796.001.0001} {\emph {\bibinfo {title}
  {Electrons and phonons: the theory of transport phenomena in solids}}}\
  (\bibinfo  {publisher} {Oxford University Press},\ \bibinfo {year}
  {1960})\BibitemShut {NoStop}%
\bibitem [{\citenamefont {Parlinski}\ \emph {et~al.}(1997)\citenamefont
  {Parlinski}, \citenamefont {Li},\ and\ \citenamefont
  {Kawazoe}}]{Parlinski-phonon-1997}%
  \BibitemOpen
  \bibfield  {author} {\bibinfo {author} {\bibfnamefont {K.}~\bibnamefont
  {Parlinski}}, \bibinfo {author} {\bibfnamefont {Z.~Q.}\ \bibnamefont {Li}}, \
  and\ \bibinfo {author} {\bibfnamefont {Y.}~\bibnamefont {Kawazoe}},\
  }\bibfield  {title} {\enquote {\bibinfo {title} {First-principles
  determination of the soft mode in cubic ${\mathrm{zro}}_{2}$},}\ }\href
  {\doibase 10.1103/PhysRevLett.78.4063} {\bibfield  {journal} {\bibinfo
  {journal} {Phys. Rev. Lett.}\ }\textbf {\bibinfo {volume} {78}},\ \bibinfo
  {pages} {4063--4066} (\bibinfo {year} {1997})}\BibitemShut {NoStop}%
\bibitem [{\citenamefont {Hellman}\ \emph {et~al.}(2013)\citenamefont
  {Hellman}, \citenamefont {Steneteg}, \citenamefont {Abrikosov},\ and\
  \citenamefont {Simak}}]{Hellman-TDEP-2013}%
  \BibitemOpen
  \bibfield  {author} {\bibinfo {author} {\bibfnamefont {O.}~\bibnamefont
  {Hellman}}, \bibinfo {author} {\bibfnamefont {P.}~\bibnamefont {Steneteg}},
  \bibinfo {author} {\bibfnamefont {I.~A.}\ \bibnamefont {Abrikosov}}, \ and\
  \bibinfo {author} {\bibfnamefont {S.~I.}\ \bibnamefont {Simak}},\ }\bibfield
  {title} {\enquote {\bibinfo {title} {Temperature dependent effective
  potential method for accurate free energy calculations of solids},}\ }\href
  {\doibase 10.1103/PhysRevB.87.104111} {\bibfield  {journal} {\bibinfo
  {journal} {Phys. Rev. B}\ }\textbf {\bibinfo {volume} {87}},\ \bibinfo
  {pages} {104111} (\bibinfo {year} {2013})}\BibitemShut {NoStop}%
\bibitem [{\citenamefont {Tadano}\ and\ \citenamefont
  {Tsuneyuki}(2018)}]{Tadano-ALM-2018}%
  \BibitemOpen
  \bibfield  {author} {\bibinfo {author} {\bibfnamefont {Terumasa}\
  \bibnamefont {Tadano}}\ and\ \bibinfo {author} {\bibfnamefont {Shinji}\
  \bibnamefont {Tsuneyuki}},\ }\bibfield  {title} {\enquote {\bibinfo {title}
  {First-principles lattice dynamics method for strongly anharmonic
  crystals},}\ }\href {\doibase 10.7566/JPSJ.87.041015} {\bibfield  {journal}
  {\bibinfo  {journal} {J. Phys. Soc. Jpn.}\ }\textbf {\bibinfo {volume}
  {87}},\ \bibinfo {pages} {041015} (\bibinfo {year} {2018})}\BibitemShut
  {NoStop}%
\bibitem [{\citenamefont {Eriksson}\ \emph {et~al.}(2019)\citenamefont
  {Eriksson}, \citenamefont {Fransson},\ and\ \citenamefont
  {Erhart}}]{hiPhive}%
  \BibitemOpen
  \bibfield  {author} {\bibinfo {author} {\bibfnamefont {F.}~\bibnamefont
  {Eriksson}}, \bibinfo {author} {\bibfnamefont {E.}~\bibnamefont {Fransson}},
  \ and\ \bibinfo {author} {\bibfnamefont {P.}~\bibnamefont {Erhart}},\
  }\bibfield  {title} {\enquote {\bibinfo {title} {The hiphive package for the
  extraction of high-order force constants by machine learning},}\ }\href
  {\doibase https://doi.org/10.1002/adts.201800184} {\bibfield  {journal}
  {\bibinfo  {journal} {Adv. Theory Simul.}\ }\textbf {\bibinfo {volume} {2}},\
  \bibinfo {pages} {1800184} (\bibinfo {year} {2019})}\BibitemShut {NoStop}%
\bibitem [{\citenamefont {Togo}(2023{\natexlab{a}})}]{phonopy-phono3py-JPSJ}%
  \BibitemOpen
  \bibfield  {author} {\bibinfo {author} {\bibfnamefont {Atsushi}\ \bibnamefont
  {Togo}},\ }\bibfield  {title} {\enquote {\bibinfo {title} {First-principles
  phonon calculations with phonopy and phono3py},}\ }\href {\doibase
  10.7566/JPSJ.92.012001} {\bibfield  {journal} {\bibinfo  {journal} {J. Phys.
  Soc. Jpn.}\ }\textbf {\bibinfo {volume} {92}},\ \bibinfo {pages} {012001}
  (\bibinfo {year} {2023}{\natexlab{a}})}\BibitemShut {NoStop}%
\bibitem [{\citenamefont {Errea}\ \emph {et~al.}(2013)\citenamefont {Errea},
  \citenamefont {Calandra},\ and\ \citenamefont {Mauri}}]{Errea-SSCHA-2013}%
  \BibitemOpen
  \bibfield  {author} {\bibinfo {author} {\bibfnamefont {Ion}\ \bibnamefont
  {Errea}}, \bibinfo {author} {\bibfnamefont {Matteo}\ \bibnamefont
  {Calandra}}, \ and\ \bibinfo {author} {\bibfnamefont {Francesco}\
  \bibnamefont {Mauri}},\ }\bibfield  {title} {\enquote {\bibinfo {title}
  {First-principles theory of anharmonicity and the inverse isotope effect in
  superconducting palladium-hydride compounds},}\ }\href {\doibase
  10.1103/PhysRevLett.111.177002} {\bibfield  {journal} {\bibinfo  {journal}
  {Phys. Rev. Lett.}\ }\textbf {\bibinfo {volume} {111}},\ \bibinfo {pages}
  {177002} (\bibinfo {year} {2013})}\BibitemShut {NoStop}%
\bibitem [{\citenamefont {{van Roekeghem}}\ \emph {et~al.}(2021)\citenamefont
  {{van Roekeghem}}, \citenamefont {Carrete},\ and\ \citenamefont
  {Mingo}}]{van-Roekeghem-2020}%
  \BibitemOpen
  \bibfield  {author} {\bibinfo {author} {\bibfnamefont {Ambroise}\
  \bibnamefont {{van Roekeghem}}}, \bibinfo {author} {\bibfnamefont {Jes\'us}\
  \bibnamefont {Carrete}}, \ and\ \bibinfo {author} {\bibfnamefont {Natalio}\
  \bibnamefont {Mingo}},\ }\bibfield  {title} {\enquote {\bibinfo {title}
  {Quantum self-consistent ab-initio lattice dynamics},}\ }\href {\doibase
  https://doi.org/10.1016/j.cpc.2021.107945} {\bibfield  {journal} {\bibinfo
  {journal} {Comput. Phys. Commun.}\ }\textbf {\bibinfo {volume} {263}},\
  \bibinfo {pages} {107945} (\bibinfo {year} {2021})}\BibitemShut {NoStop}%
\bibitem [{\citenamefont {Nye}(1985)}]{Physical-Properties-of-Crystals}%
  \BibitemOpen
  \bibfield  {author} {\bibinfo {author} {\bibfnamefont {J.~F.}\ \bibnamefont
  {Nye}},\ }\href@noop {} {\emph {\bibinfo {title} {Physical Properties of
  Crystals}}}\ (\bibinfo  {publisher} {Oxford University Press},\ \bibinfo
  {year} {1985})\BibitemShut {NoStop}%
\bibitem [{\citenamefont {El-Batanouny}\ and\ \citenamefont
  {Wooten}(2008)}]{el-batanouny_wooten_2008}%
  \BibitemOpen
  \bibfield  {author} {\bibinfo {author} {\bibfnamefont {M.}~\bibnamefont
  {El-Batanouny}}\ and\ \bibinfo {author} {\bibfnamefont {F.}~\bibnamefont
  {Wooten}},\ }\href {\doibase 10.1017/CBO9780511755736} {\emph {\bibinfo
  {title} {Symmetry and Condensed Matter Physics: A Computational Approach}}}\
  (\bibinfo  {publisher} {Cambridge University Press},\ \bibinfo {year}
  {2008})\BibitemShut {NoStop}%
\bibitem [{\citenamefont {Venkataraman}\ \emph {et~al.}(1975)\citenamefont
  {Venkataraman}, \citenamefont {Feldkamp},\ and\ \citenamefont
  {Sahni}}]{Dynamics-of-perfect-crystals}%
  \BibitemOpen
  \bibfield  {author} {\bibinfo {author} {\bibfnamefont {G.}~\bibnamefont
  {Venkataraman}}, \bibinfo {author} {\bibfnamefont {L.~A.}\ \bibnamefont
  {Feldkamp}}, \ and\ \bibinfo {author} {\bibfnamefont {V.~C.}\ \bibnamefont
  {Sahni}},\ }\href@noop {} {\emph {\bibinfo {title} {Dynamics of perfect
  crystals}}}\ (\bibinfo  {publisher} {MIT press},\ \bibinfo {year}
  {1975})\BibitemShut {NoStop}%
\bibitem [{\citenamefont {Zhou}\ \emph {et~al.}(2014)\citenamefont {Zhou},
  \citenamefont {Nielson}, \citenamefont {Xia},\ and\ \citenamefont
  {Ozoli\ifmmode \mbox{\c{n}}\else \c{n}\fi{}\ifmmode~\check{s}\else
  \v{s}\fi{}}}]{Zhou-PRL-compressive-sensing-FC-2014}%
  \BibitemOpen
  \bibfield  {author} {\bibinfo {author} {\bibfnamefont {Fei}\ \bibnamefont
  {Zhou}}, \bibinfo {author} {\bibfnamefont {Weston}\ \bibnamefont {Nielson}},
  \bibinfo {author} {\bibfnamefont {Yi}~\bibnamefont {Xia}}, \ and\ \bibinfo
  {author} {\bibfnamefont {Vidvuds}\ \bibnamefont {Ozoli\ifmmode
  \mbox{\c{n}}\else \c{n}\fi{}\ifmmode~\check{s}\else \v{s}\fi{}}},\ }\bibfield
   {title} {\enquote {\bibinfo {title} {Lattice anharmonicity and thermal
  conductivity from compressive sensing of first-principles calculations},}\
  }\href {\doibase 10.1103/PhysRevLett.113.185501} {\bibfield  {journal}
  {\bibinfo  {journal} {Phys. Rev. Lett.}\ }\textbf {\bibinfo {volume} {113}},\
  \bibinfo {pages} {185501} (\bibinfo {year} {2014})}\BibitemShut {NoStop}%
\bibitem [{\citenamefont {Tadano}\ and\ \citenamefont
  {Tsuneyuki}(2015)}]{Tadano-2015}%
  \BibitemOpen
  \bibfield  {author} {\bibinfo {author} {\bibfnamefont {T.}~\bibnamefont
  {Tadano}}\ and\ \bibinfo {author} {\bibfnamefont {S.}~\bibnamefont
  {Tsuneyuki}},\ }\bibfield  {title} {\enquote {\bibinfo {title}
  {Self-consistent phonon calculations of lattice dynamical properties in cubic
  ${\mathrm{srtio}}_{3}$ with first-principles anharmonic force constants},}\
  }\href {\doibase 10.1103/PhysRevB.92.054301} {\bibfield  {journal} {\bibinfo
  {journal} {Phys. Rev. B}\ }\textbf {\bibinfo {volume} {92}},\ \bibinfo
  {pages} {054301} (\bibinfo {year} {2015})}\BibitemShut {NoStop}%
\bibitem [{\citenamefont {Seko}\ and\ \citenamefont
  {Togo}(2024)}]{PhysRevB.110.214302}%
  \BibitemOpen
  \bibfield  {author} {\bibinfo {author} {\bibfnamefont {Atsuto}\ \bibnamefont
  {Seko}}\ and\ \bibinfo {author} {\bibfnamefont {Atsushi}\ \bibnamefont
  {Togo}},\ }\bibfield  {title} {\enquote {\bibinfo {title} {Projector-based
  efficient estimation of force constants},}\ }\href {\doibase
  10.1103/PhysRevB.110.214302} {\bibfield  {journal} {\bibinfo  {journal}
  {Phys. Rev. B}\ }\textbf {\bibinfo {volume} {110}},\ \bibinfo {pages}
  {214302} (\bibinfo {year} {2024})}\BibitemShut {NoStop}%
\bibitem [{\citenamefont {Seko}\ and\ \citenamefont {Togo}()}]{symfc-project}%
  \BibitemOpen
  \bibfield  {author} {\bibinfo {author} {\bibfnamefont {Atsuto}\ \bibnamefont
  {Seko}}\ and\ \bibinfo {author} {\bibfnamefont {Atsushi}\ \bibnamefont
  {Togo}},\ }\href@noop {} {\enquote {\bibinfo {title} {Symfc},}\ }\bibinfo
  {howpublished} {\url{https://github.com/symfc/symfc}}\BibitemShut {NoStop}%
\bibitem [{\citenamefont {Togo}\ and\ \citenamefont
  {Seko}(2024)}]{10.1063/5.0211296}%
  \BibitemOpen
  \bibfield  {author} {\bibinfo {author} {\bibfnamefont {Atsushi}\ \bibnamefont
  {Togo}}\ and\ \bibinfo {author} {\bibfnamefont {Atsuto}\ \bibnamefont
  {Seko}},\ }\bibfield  {title} {\enquote {\bibinfo {title} {{On-the-fly
  training of polynomial machine learning potentials in computing lattice
  thermal conductivity}},}\ }\href {\doibase 10.1063/5.0211296} {\bibfield
  {journal} {\bibinfo  {journal} {J. Chem. Phys.}\ }\textbf {\bibinfo {volume}
  {160}},\ \bibinfo {pages} {211001} (\bibinfo {year} {2024})}\BibitemShut
  {NoStop}%
\bibitem [{\citenamefont {Wakai}\ \emph {et~al.}(2026)\citenamefont {Wakai},
  \citenamefont {Seko},\ and\ \citenamefont
  {Tanaka}}]{wakai2026globalstructuresearchesvarying}%
  \BibitemOpen
  \bibfield  {author} {\bibinfo {author} {\bibfnamefont {H}~\bibnamefont
  {Wakai}}, \bibinfo {author} {\bibfnamefont {A}~\bibnamefont {Seko}}, \ and\
  \bibinfo {author} {\bibfnamefont {I}~\bibnamefont {Tanaka}},\ }\href
  {https://arxiv.org/abs/2503.22596} {\enquote {\bibinfo {title} {Global
  structure searches under varying temperatures and pressures using polynomial
  machine learning potentials: A case study on silicon},}\ } (\bibinfo {year}
  {2026}),\ \Eprint {http://arxiv.org/abs/2503.22596} {arXiv:2503.22596
  [cond-mat.mtrl-sci]} \BibitemShut {NoStop}%
\bibitem [{\citenamefont {von Neumann}(1949)}]{Neumann1949OnRO}%
  \BibitemOpen
  \bibfield  {author} {\bibinfo {author} {\bibfnamefont {John}\ \bibnamefont
  {von Neumann}},\ }\bibfield  {title} {\enquote {\bibinfo {title} {On rings of
  operators. reduction theory},}\ }\href
  {https://api.semanticscholar.org/CorpusID:124439084} {\bibfield  {journal}
  {\bibinfo  {journal} {Annals of Math.}\ }\textbf {\bibinfo {volume} {50}},\
  \bibinfo {pages} {401} (\bibinfo {year} {1949})}\BibitemShut {NoStop}%
\bibitem [{\citenamefont {Halperin}(1962)}]{halperin1962product}%
  \BibitemOpen
  \bibfield  {author} {\bibinfo {author} {\bibfnamefont {Israel}\ \bibnamefont
  {Halperin}},\ }\bibfield  {title} {\enquote {\bibinfo {title} {The product of
  projection operators},}\ }\href@noop {} {\bibfield  {journal} {\bibinfo
  {journal} {Acta Sci. Math.(Szeged)}\ }\textbf {\bibinfo {volume} {23}},\
  \bibinfo {pages} {96--99} (\bibinfo {year} {1962})}\BibitemShut {NoStop}%
\bibitem [{\citenamefont {Piziak}\ \emph {et~al.}(1999)\citenamefont {Piziak},
  \citenamefont {Odell},\ and\ \citenamefont {Hahn}}]{PIZIAK199967}%
  \BibitemOpen
  \bibfield  {author} {\bibinfo {author} {\bibfnamefont {R.}~\bibnamefont
  {Piziak}}, \bibinfo {author} {\bibfnamefont {P.L.}\ \bibnamefont {Odell}}, \
  and\ \bibinfo {author} {\bibfnamefont {R.}~\bibnamefont {Hahn}},\ }\bibfield
  {title} {\enquote {\bibinfo {title} {Constructing projections on sums and
  intersections},}\ }\href {\doibase
  https://doi.org/10.1016/S0898-1221(98)00242-9} {\bibfield  {journal}
  {\bibinfo  {journal} {Comput. Math. Appl.}\ }\textbf {\bibinfo {volume}
  {37}},\ \bibinfo {pages} {67--74} (\bibinfo {year} {1999})}\BibitemShut
  {NoStop}%
\bibitem [{\citenamefont {Gal{\'a}ntai}(2013)}]{galantai2013projectors}%
  \BibitemOpen
  \bibfield  {author} {\bibinfo {author} {\bibfnamefont {Aur{\'e}l}\
  \bibnamefont {Gal{\'a}ntai}},\ }\href@noop {} {\emph {\bibinfo {title}
  {Projectors and projection methods}}},\ Vol.~\bibinfo {volume} {6}\ (\bibinfo
   {publisher} {Springer Science \& Business Media},\ \bibinfo {year}
  {2013})\BibitemShut {NoStop}%
\bibitem [{\citenamefont {Nishiyama}\ \emph {et~al.}(2020)\citenamefont
  {Nishiyama}, \citenamefont {Seko},\ and\ \citenamefont
  {Tanaka}}]{PhysRevMaterials.4.123607}%
  \BibitemOpen
  \bibfield  {author} {\bibinfo {author} {\bibfnamefont {Takayuki}\
  \bibnamefont {Nishiyama}}, \bibinfo {author} {\bibfnamefont {Atsuto}\
  \bibnamefont {Seko}}, \ and\ \bibinfo {author} {\bibfnamefont {Isao}\
  \bibnamefont {Tanaka}},\ }\bibfield  {title} {\enquote {\bibinfo {title}
  {Application of machine learning potentials to predict grain boundary
  properties in fcc elemental metals},}\ }\href {\doibase
  10.1103/PhysRevMaterials.4.123607} {\bibfield  {journal} {\bibinfo  {journal}
  {Phys. Rev. Mater.}\ }\textbf {\bibinfo {volume} {4}},\ \bibinfo {pages}
  {123607} (\bibinfo {year} {2020})}\BibitemShut {NoStop}%
\bibitem [{\citenamefont {Seko}(2023)}]{doi:10.1063/5.0129045}%
  \BibitemOpen
  \bibfield  {author} {\bibinfo {author} {\bibfnamefont {Atsuto}\ \bibnamefont
  {Seko}},\ }\bibfield  {title} {\enquote {\bibinfo {title} {Tutorial:
  Systematic development of polynomial machine learning potentials for
  elemental and alloy systems},}\ }\href {\doibase 10.1063/5.0129045}
  {\bibfield  {journal} {\bibinfo  {journal} {J. Appl. Phys.}\ }\textbf
  {\bibinfo {volume} {133}},\ \bibinfo {pages} {011101} (\bibinfo {year}
  {2023})}\BibitemShut {NoStop}%
\bibitem [{\citenamefont {Togo}\ and\ \citenamefont
  {Tanaka}(2015)}]{Togo20151}%
  \BibitemOpen
  \bibfield  {author} {\bibinfo {author} {\bibfnamefont {Atsushi}\ \bibnamefont
  {Togo}}\ and\ \bibinfo {author} {\bibfnamefont {Isao}\ \bibnamefont
  {Tanaka}},\ }\bibfield  {title} {\enquote {\bibinfo {title} {First principles
  phonon calculations in materials science},}\ }\href {\doibase
  https://doi.org/10.1016/j.scriptamat.2015.07.021} {\bibfield  {journal}
  {\bibinfo  {journal} {Script. Mater.}\ }\textbf {\bibinfo {volume} {108}},\
  \bibinfo {pages} {1 -- 5} (\bibinfo {year} {2015})}\BibitemShut {NoStop}%
\bibitem [{\citenamefont {Togo}\ \emph {et~al.}(2023)\citenamefont {Togo},
  \citenamefont {Chaput}, \citenamefont {Tadano},\ and\ \citenamefont
  {Tanaka}}]{phonopy-phono3py-JPCM}%
  \BibitemOpen
  \bibfield  {author} {\bibinfo {author} {\bibfnamefont {Atsushi}\ \bibnamefont
  {Togo}}, \bibinfo {author} {\bibfnamefont {Laurent}\ \bibnamefont {Chaput}},
  \bibinfo {author} {\bibfnamefont {Terumasa}\ \bibnamefont {Tadano}}, \ and\
  \bibinfo {author} {\bibfnamefont {Isao}\ \bibnamefont {Tanaka}},\ }\bibfield
  {title} {\enquote {\bibinfo {title} {Implementation strategies in phonopy and
  phono3py},}\ }\href {\doibase 10.1088/1361-648X/acd831} {\bibfield  {journal}
  {\bibinfo  {journal} {J. Phys. Condens. Matter}\ }\textbf {\bibinfo {volume}
  {35}},\ \bibinfo {pages} {353001} (\bibinfo {year} {2023})}\BibitemShut
  {NoStop}%
\bibitem [{\citenamefont {Seko}\ \emph {et~al.}(2019)\citenamefont {Seko},
  \citenamefont {Togo},\ and\ \citenamefont {Tanaka}}]{PhysRevB.99.214108}%
  \BibitemOpen
  \bibfield  {author} {\bibinfo {author} {\bibfnamefont {Atsuto}\ \bibnamefont
  {Seko}}, \bibinfo {author} {\bibfnamefont {Atsushi}\ \bibnamefont {Togo}}, \
  and\ \bibinfo {author} {\bibfnamefont {Isao}\ \bibnamefont {Tanaka}},\
  }\bibfield  {title} {\enquote {\bibinfo {title} {Group-theoretical high-order
  rotational invariants for structural representations: Application to
  linearized machine learning interatomic potential},}\ }\href {\doibase
  10.1103/PhysRevB.99.214108} {\bibfield  {journal} {\bibinfo  {journal} {Phys.
  Rev. B}\ }\textbf {\bibinfo {volume} {99}},\ \bibinfo {pages} {214108}
  (\bibinfo {year} {2019})}\BibitemShut {NoStop}%
\bibitem [{Mac()}]{MachineLearningPotentialRepository}%
  \BibitemOpen
  \href {https://sekocha.github.io} {}\bibinfo {note} {{A. Seko}, {Polynomial}
  {Machine} {Learning} {Potential} {Repository} at {Kyoto} {University},
  \url{https://sekocha.github.io}}\BibitemShut {NoStop}%
\bibitem [{\citenamefont {Togo}(2023{\natexlab{b}})}]{Togo2023MDRPhonon}%
  \BibitemOpen
  \bibfield  {author} {\bibinfo {author} {\bibfnamefont {Atsushi}\ \bibnamefont
  {Togo}},\ }\href {\doibase 10.48505/nims.4197} {\enquote {\bibinfo {title}
  {Phonondb},}\ } (\bibinfo {year} {2023}{\natexlab{b}})\BibitemShut {NoStop}%
\bibitem [{\citenamefont {Bl{\"o}chl}(1994)}]{PAW1}%
  \BibitemOpen
  \bibfield  {author} {\bibinfo {author} {\bibfnamefont {P~E}\ \bibnamefont
  {Bl{\"o}chl}},\ }\bibfield  {title} {\enquote {\bibinfo {title} {Projector
  augmented-wave method},}\ }\href@noop {} {\bibfield  {journal} {\bibinfo
  {journal} {Phys. Rev. B}\ }\textbf {\bibinfo {volume} {50}},\ \bibinfo
  {pages} {17953--17979} (\bibinfo {year} {1994})}\BibitemShut {NoStop}%
\bibitem [{\citenamefont {Perdew}\ \emph {et~al.}(1996)\citenamefont {Perdew},
  \citenamefont {Burke},\ and\ \citenamefont {Ernzerhof}}]{GGA:PBE96}%
  \BibitemOpen
  \bibfield  {author} {\bibinfo {author} {\bibfnamefont {J~P}\ \bibnamefont
  {Perdew}}, \bibinfo {author} {\bibfnamefont {K}~\bibnamefont {Burke}}, \ and\
  \bibinfo {author} {\bibfnamefont {M}~\bibnamefont {Ernzerhof}},\ }\bibfield
  {title} {\enquote {\bibinfo {title} {Generalized gradient approximation made
  simple},}\ }\href@noop {} {\bibfield  {journal} {\bibinfo  {journal} {Phys.
  Rev. Lett.}\ }\textbf {\bibinfo {volume} {77}},\ \bibinfo {pages}
  {3865--3868} (\bibinfo {year} {1996})}\BibitemShut {NoStop}%
\bibitem [{\citenamefont {Kresse}\ and\ \citenamefont {Hafner}(1993)}]{VASP1}%
  \BibitemOpen
  \bibfield  {author} {\bibinfo {author} {\bibfnamefont {G}~\bibnamefont
  {Kresse}}\ and\ \bibinfo {author} {\bibfnamefont {J}~\bibnamefont {Hafner}},\
  }\bibfield  {title} {\enquote {\bibinfo {title} {{$Ab$} $initio$
  molecular-dynamics for liquid-metals},}\ }\href@noop {} {\bibfield  {journal}
  {\bibinfo  {journal} {Phys. Rev. B}\ }\textbf {\bibinfo {volume} {47}},\
  \bibinfo {pages} {558--561} (\bibinfo {year} {1993})}\BibitemShut {NoStop}%
\bibitem [{\citenamefont {Kresse}\ and\ \citenamefont
  {Furthm{\"u}ller}(1996)}]{VASP2}%
  \BibitemOpen
  \bibfield  {author} {\bibinfo {author} {\bibfnamefont {G}~\bibnamefont
  {Kresse}}\ and\ \bibinfo {author} {\bibfnamefont {J}~\bibnamefont
  {Furthm{\"u}ller}},\ }\bibfield  {title} {\enquote {\bibinfo {title}
  {Efficient iterative schemes for $ab$ $initio$ total-energy calculations
  using a plane-wave basis set},}\ }\href@noop {} {\bibfield  {journal}
  {\bibinfo  {journal} {Phys. Rev. B}\ }\textbf {\bibinfo {volume} {54}},\
  \bibinfo {pages} {11169--11186} (\bibinfo {year} {1996})}\BibitemShut
  {NoStop}%
\bibitem [{\citenamefont {Kresse}\ and\ \citenamefont {Joubert}(1999)}]{PAW2}%
  \BibitemOpen
  \bibfield  {author} {\bibinfo {author} {\bibfnamefont {G}~\bibnamefont
  {Kresse}}\ and\ \bibinfo {author} {\bibfnamefont {D}~\bibnamefont
  {Joubert}},\ }\bibfield  {title} {\enquote {\bibinfo {title} {From ultrasoft
  pseudopotentials to the projector augmented-wave method},}\ }\href@noop {}
  {\bibfield  {journal} {\bibinfo  {journal} {Phys. Rev. B}\ }\textbf {\bibinfo
  {volume} {59}},\ \bibinfo {pages} {1758--1775} (\bibinfo {year}
  {1999})}\BibitemShut {NoStop}%
\bibitem [{\citenamefont {Togo}\ \emph {et~al.}(2015)\citenamefont {Togo},
  \citenamefont {Chaput},\ and\ \citenamefont {Tanaka}}]{phono3py}%
  \BibitemOpen
  \bibfield  {author} {\bibinfo {author} {\bibfnamefont {A.}~\bibnamefont
  {Togo}}, \bibinfo {author} {\bibfnamefont {L.}~\bibnamefont {Chaput}}, \ and\
  \bibinfo {author} {\bibfnamefont {I.}~\bibnamefont {Tanaka}},\ }\bibfield
  {title} {\enquote {\bibinfo {title} {Distributions of phonon lifetimes in
  brillouin zones},}\ }\href@noop {} {\bibfield  {journal} {\bibinfo  {journal}
  {Phys. Rev. B}\ }\textbf {\bibinfo {volume} {91}},\ \bibinfo {pages} {094306}
  (\bibinfo {year} {2015})}\BibitemShut {NoStop}%
\bibitem [{\citenamefont {Peierls}(1929)}]{Peierls-Boltzmann-1929}%
  \BibitemOpen
  \bibfield  {author} {\bibinfo {author} {\bibfnamefont {R.~E.}\ \bibnamefont
  {Peierls}},\ }\bibfield  {title} {\enquote {\bibinfo {title} {Zur kinetischen
  theorie der wärmeleitung in kristallen},}\ }\href {\doibase
  https://doi.org/10.1002/andp.19293950803} {\bibfield  {journal} {\bibinfo
  {journal} {Ann. Phys.}\ }\textbf {\bibinfo {volume} {395}},\ \bibinfo {pages}
  {1055--1101} (\bibinfo {year} {1929})}\BibitemShut {NoStop}%
\bibitem [{\citenamefont {Peierls}(2001)}]{Peierls-Quantum-Theory-of-Solids}%
  \BibitemOpen
  \bibfield  {author} {\bibinfo {author} {\bibfnamefont {R.~E.}\ \bibnamefont
  {Peierls}},\ }\href@noop {} {\emph {\bibinfo {title} {Quantum theory of
  solids}}}\ (\bibinfo  {publisher} {Oxford University Press},\ \bibinfo {year}
  {2001})\BibitemShut {NoStop}%
\bibitem [{\citenamefont {Allen}\ and\ \citenamefont
  {Perebeinos}(2018)}]{Allen-LTC-2018}%
  \BibitemOpen
  \bibfield  {author} {\bibinfo {author} {\bibfnamefont {Philip~B.}\
  \bibnamefont {Allen}}\ and\ \bibinfo {author} {\bibfnamefont {Vasili}\
  \bibnamefont {Perebeinos}},\ }\bibfield  {title} {\enquote {\bibinfo {title}
  {Temperature in a peierls-boltzmann treatment of nonlocal phonon heat
  transport},}\ }\href {\doibase 10.1103/PhysRevB.98.085427} {\bibfield
  {journal} {\bibinfo  {journal} {Phys. Rev. B}\ }\textbf {\bibinfo {volume}
  {98}},\ \bibinfo {pages} {085427} (\bibinfo {year} {2018})}\BibitemShut
  {NoStop}%
\bibitem [{\citenamefont {Fujii}\ and\ \citenamefont
  {Seko}(2022)}]{FUJII2022111137}%
  \BibitemOpen
  \bibfield  {author} {\bibinfo {author} {\bibfnamefont {Susumu}\ \bibnamefont
  {Fujii}}\ and\ \bibinfo {author} {\bibfnamefont {Atsuto}\ \bibnamefont
  {Seko}},\ }\bibfield  {title} {\enquote {\bibinfo {title} {Structure and
  lattice thermal conductivity of grain boundaries in silicon by using machine
  learning potential and molecular dynamics},}\ }\href {\doibase
  https://doi.org/10.1016/j.commatsci.2021.111137} {\bibfield  {journal}
  {\bibinfo  {journal} {Comput. Mater. Sci.}\ }\textbf {\bibinfo {volume}
  {204}},\ \bibinfo {pages} {111137} (\bibinfo {year} {2022})}\BibitemShut
  {NoStop}%
\bibitem [{\citenamefont {Pyle}(1967)}]{pyle1967generalized}%
  \BibitemOpen
  \bibfield  {author} {\bibinfo {author} {\bibfnamefont {L.~Duane}\
  \bibnamefont {Pyle}},\ }\bibfield  {title} {\enquote {\bibinfo {title} {A
  generalized inverse $\epsilon$-algorithm for constructing intersection
  projection matrices, with applications},}\ }\href {\doibase
  10.1007/BF02165164} {\bibfield  {journal} {\bibinfo  {journal} {Numerische
  Mathematik}\ }\textbf {\bibinfo {volume} {10}},\ \bibinfo {pages} {86--102}
  (\bibinfo {year} {1967})}\BibitemShut {NoStop}%
\end{thebibliography}%

\end{document}